\documentclass[aps,prl,twocolumn,superscriptaddress]{revtex4}

\usepackage{graphicx,latexsym,color}

\begin{document}


\title{Power Law Distributions of Patents as Indicators of Innovation.}

\author{D. R. J. O'Neale}
\email[]{d.oneale@irl.cri.nz}
\affiliation{Industrial Research Ltd, Lower Hutt 5040, New Zealand}
\author{S. C. Hendy}
\affiliation{Industrial Research Ltd, Lower Hutt 5040, New Zealand}
\affiliation{MacDiarmid Institute for Advanced Materials
and Nanotechnology, School of Chemical and Physical Sciences,
Victoria University of Wellington, Wellington 6140, New Zealand}


\date{\today}

\begin{abstract}
The total number of patents produced by a country (or the number of patents produced per capita) is often used as an indicator for innovation.  Here we present evidence that the distribution of patents amongst applicants within many OECD countries is well-described by power laws with exponents that vary between 1.66 (Japan) and 2.37 (Poland).  Using simulations based on simple preferential attachment-type rules that generate power laws, we find we can explain some of the variation in exponents between countries, with countries that have larger numbers of patents per applicant generally exhibiting smaller exponents in both the simulated and actual data.  Similarly we find that the exponents for most countries are inversely correlated with other indicators of innovation, such as R\&D intensity or the ubiquity of export baskets.  This suggests that in more advanced economies, which tend to have smaller values of the exponent, a greater proportion of the total number of patents are filed by large companies than in less advanced countries. 
\end{abstract}


\maketitle


Despite the crucial role that firms and their inter-relationships play in the performance of national economies, a complete description of this complex system has eluded economists.  More than a century ago, one of the fathers of modern economics, Alfred Marshall, drew on an analogy with forest ecosystems to describe this system: {\it ``\ldots we may read a lesson from the young trees of the forest as they struggle upwards through the benumbing shade of their older rivals.  Many succumb on the way, and only a few survive\ldots And as with the growth of trees, so it was with the growth of businesses\ldots''} \cite{Marshall:1920}.  Today we have indications that this may be more than just a metaphor.  Many observations support the idea that, as with the distribution of biomass and metabolic rates amongst biological organisms \cite{West:1997}, the distribution of firm sizes follows a power law \cite{Axtell:2001, Fujiwara:2004, Okuyama:1999, Luttmer:2007, Luttmer:2011}.  

\begin{figure}[b]
		\includegraphics[width=0.80\columnwidth]{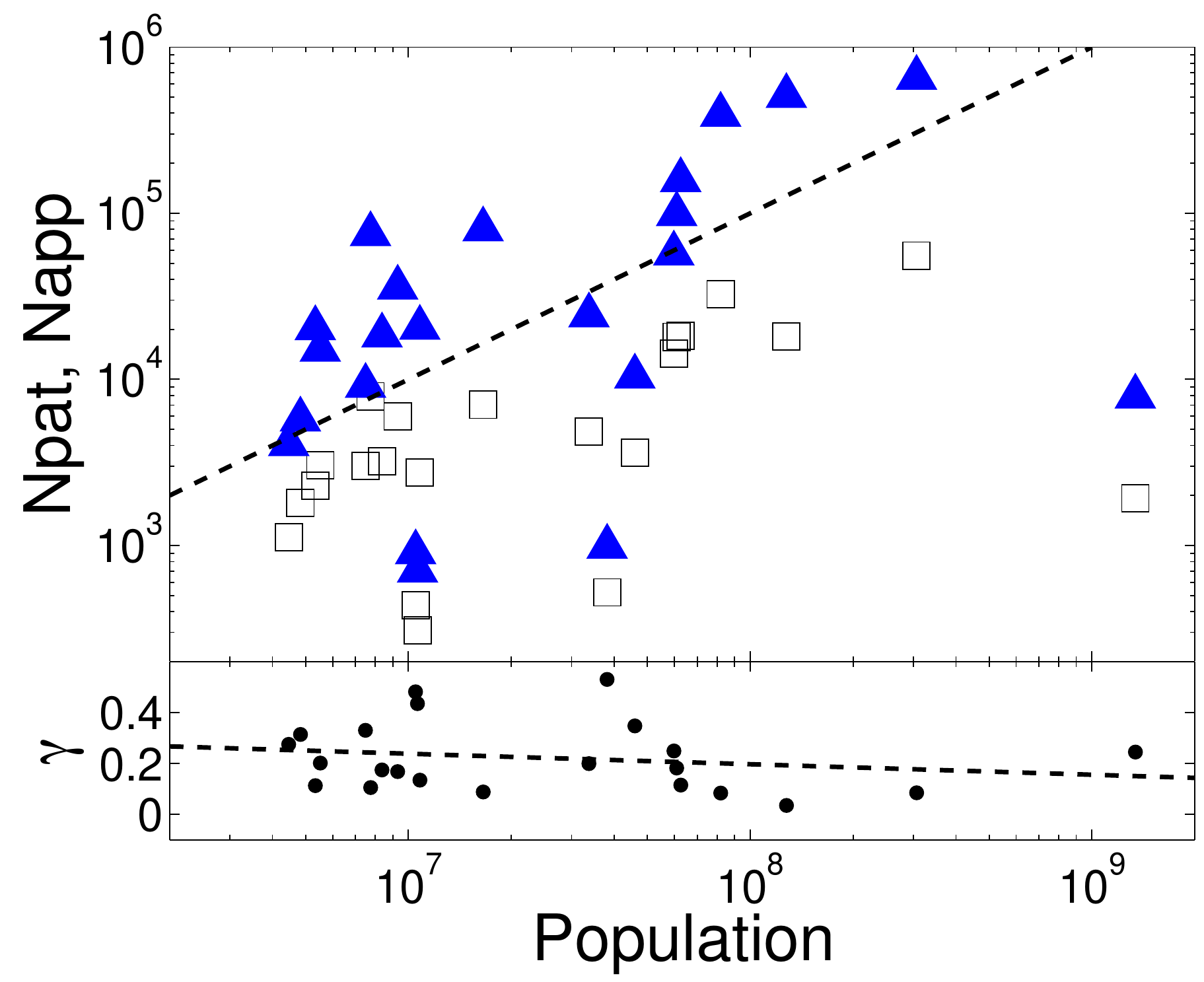}
		\caption{\label{fig1}(Color online) Upper plot: Number of EPO patent applications (filled triangles) and unique applicants (open squares) versus national population, for the 22 countries in the OECD HAN data set. The dashed line indicates the slope which the data would follow if they scaled linearly --- in the absence of agglomeration effects. Lower plot: Ratio of number of applicants to number of patents ($\gamma=N_{app}/N_{pat}$) for the same data. The least squares best fit has a slope of $-0.041$ and $R^2=0.038$ indicating a poor correlation and little dependence on population.}
\end{figure}

Power laws seem to abound in economic data. When aggregated at the city level, quantities such as the number of new patents, inventors, R\&D establishments and even productivity have been found to follow power law scaling with respect to city size \cite{Bettencourt:2007, Hendy:2011}. Crucially, these measures of innovation scale super-linearly; larger cities are more productive per capita than smaller cities \cite{Bettencourt:2010}. However when patent data is aggregated at the level of countries, rather than cities or regions, the picture is entirely different. In contrast to the findings in \cite{Bettencourt:2007, Bettencourt:2010, Hendy:2011} where data are tightly clustered and show clear super-linear scaling, the number of patents varies roughly linearly with a country's population and is poorly correlated in comparison with the results for cities, Fig.~\ref{fig1}.  Economic geography tells us that cities exist to exploit the benefits of agglomeration, so it is not necessarily surprising that super-linear scaling for cities can be seen in patent data. Countries exist for more complicated reasons. 
 
In this article we investigate how intellectual property is distributed in national economic ecosystems.  In particular, we consider the distribution of patents among applicants within countries and find compelling evidence that this distribution follows a power law in many instances.  However these power laws are not universal: the best-fit exponents for these distributions differs from country to country by a statistically significant amount.  This suggests that firms within the ecosystems of different countries experience different environments which influence their patenting behavior.  Using a simple preferential attachment model, based on the Yule process, we show it is possible to reproduce the qualitative features observed in the empirical data and to explain some of the variation of exponents between countries.  We also find that the value of the power law exponent is inversely correlated with a number of indicators that are commonly linked with innovation, such as research and development (R\&D) intensity.  Interestingly we find that the value of the exponent saturates at high R\&D intensities.  

\begin{figure*}[tb]
\includegraphics[width=0.16\textwidth]{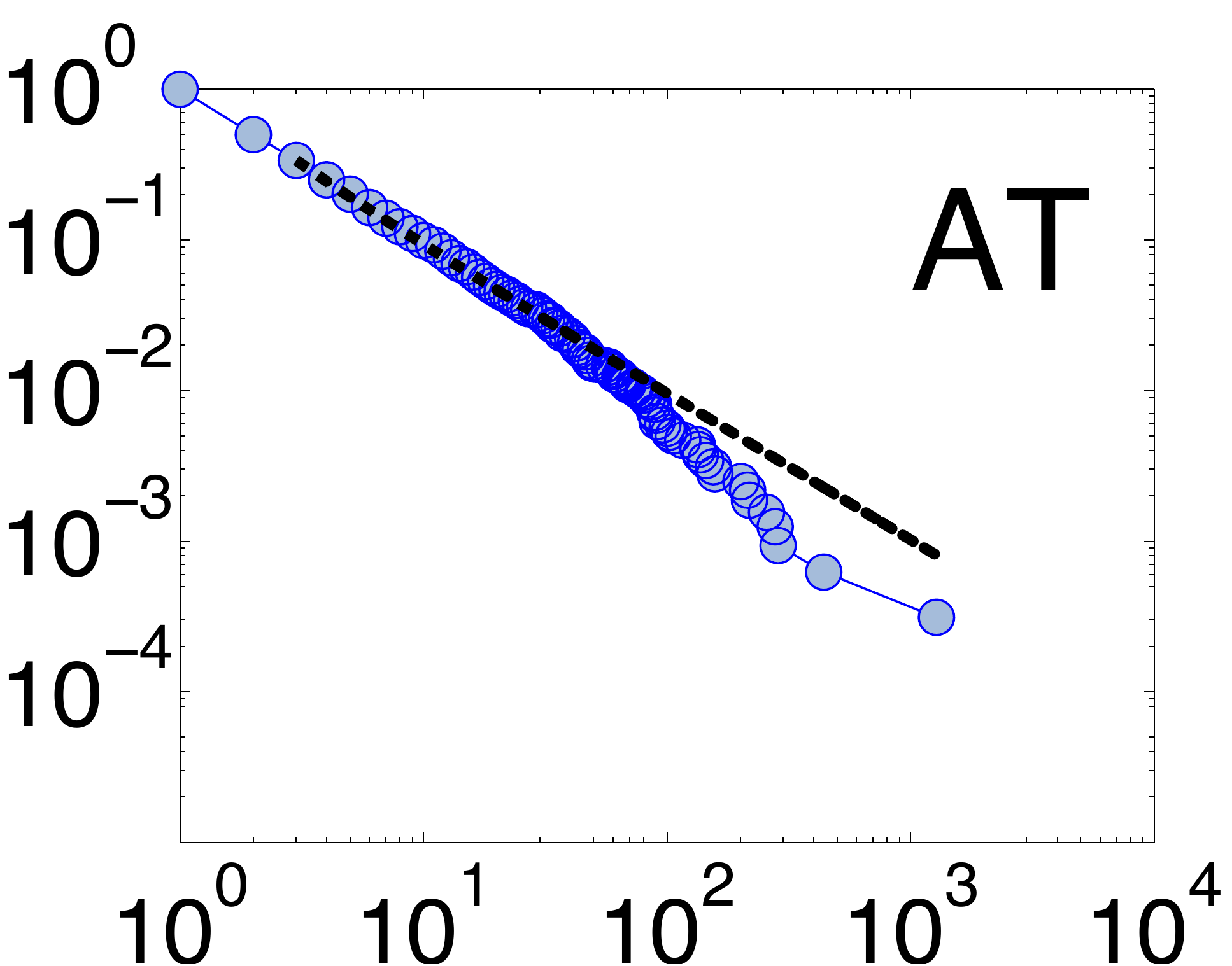}
\includegraphics[width=0.16\textwidth]{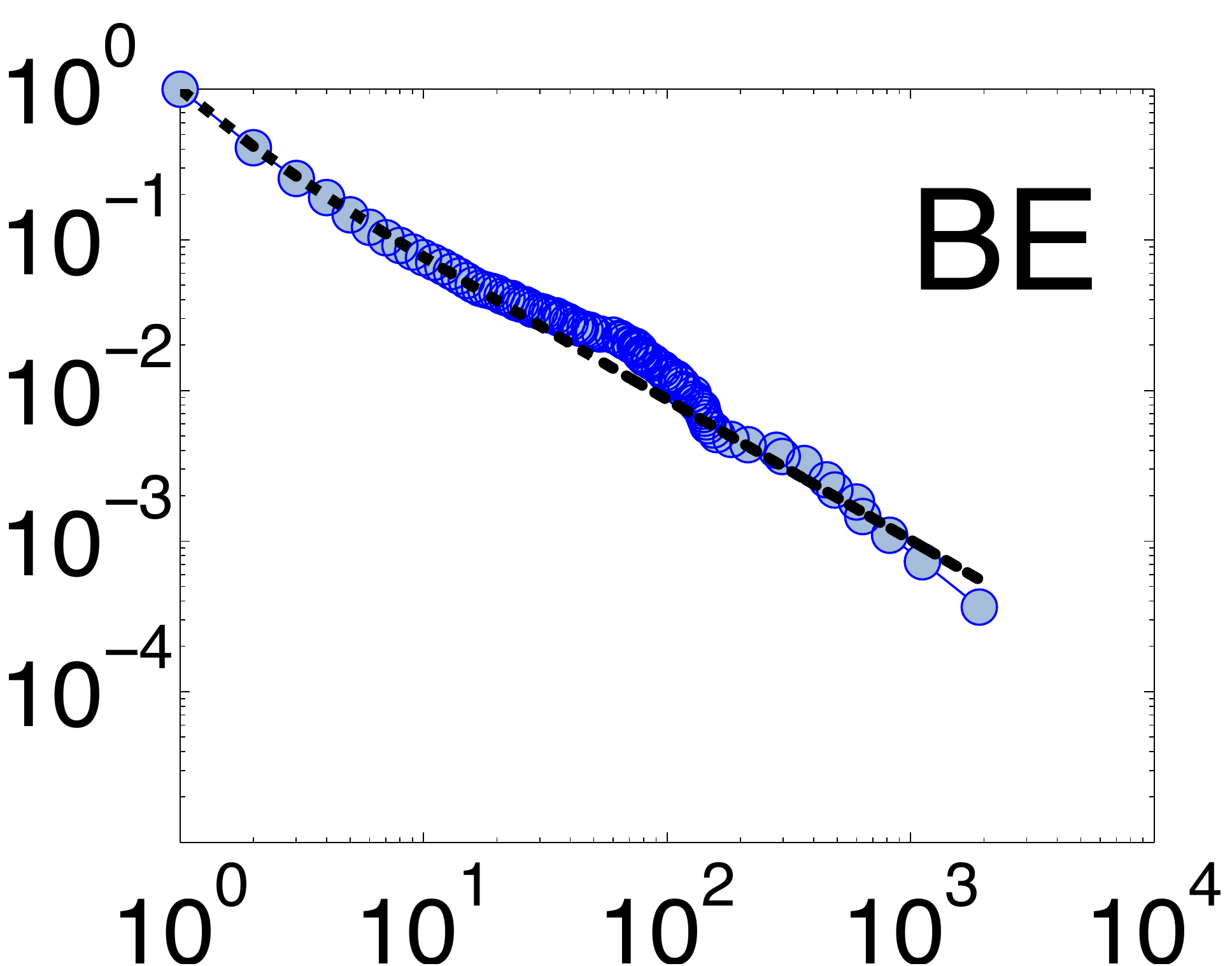}
\includegraphics[width=0.16\textwidth]{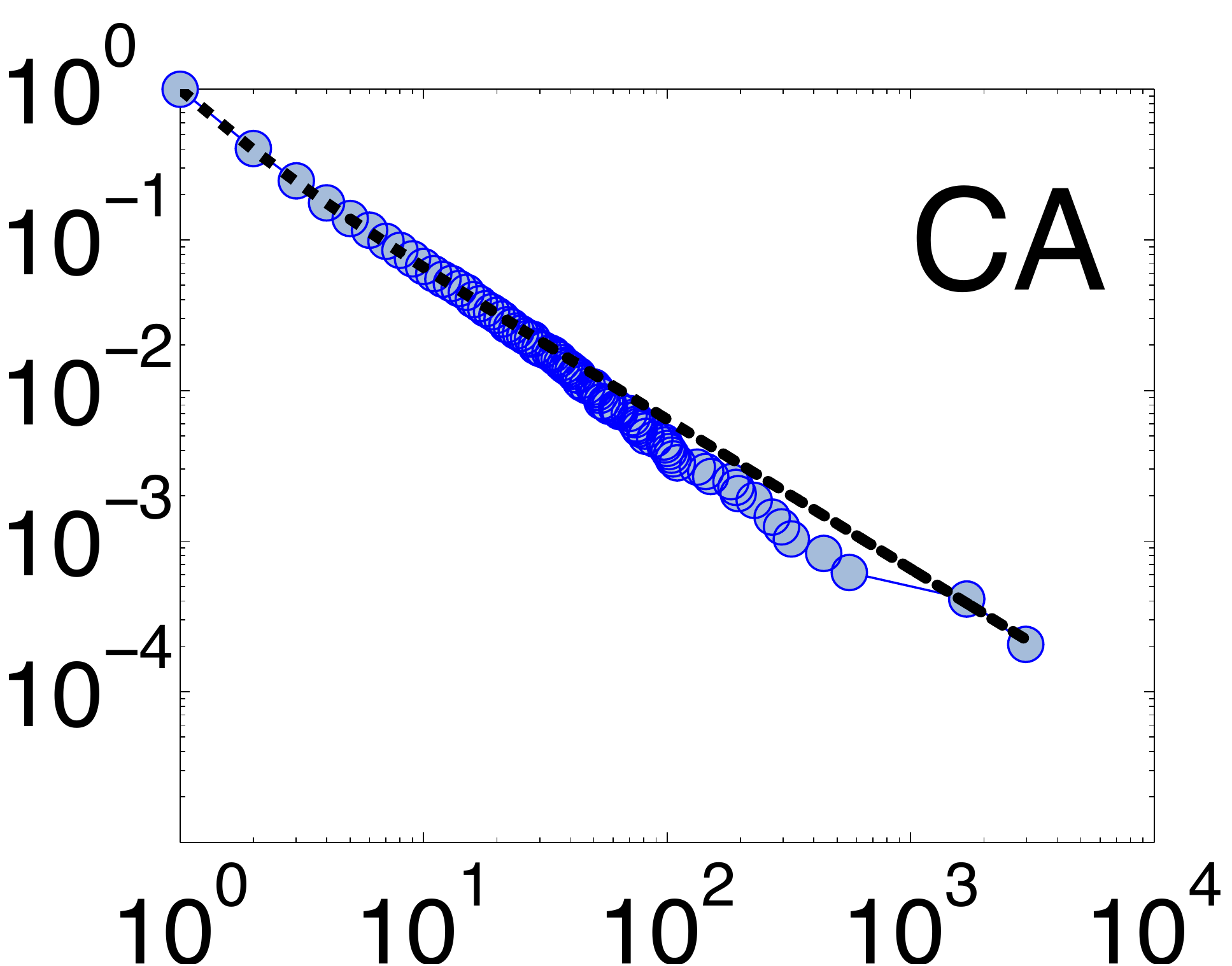}
\includegraphics[width=0.16\textwidth]{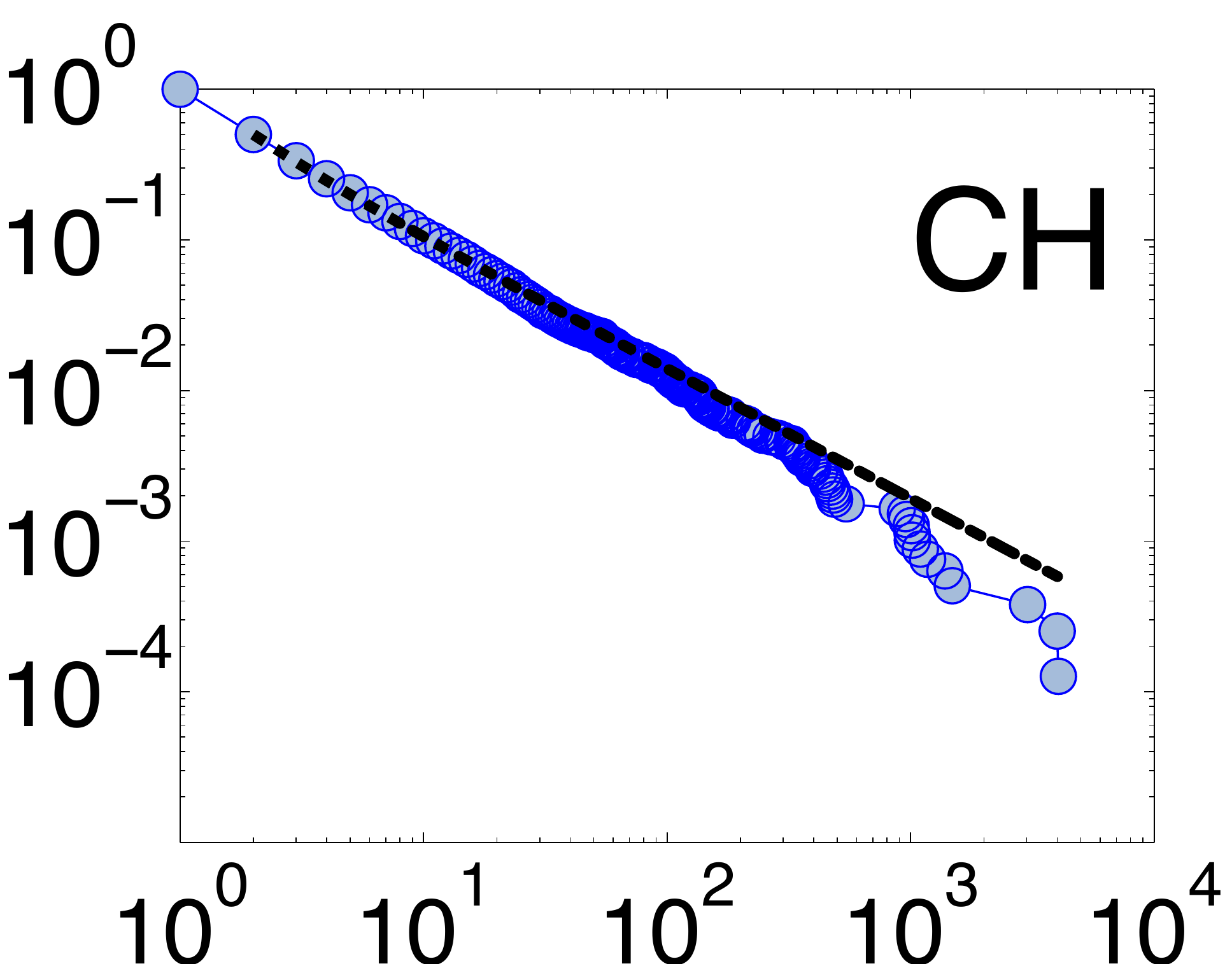}
\includegraphics[width=0.16\textwidth]{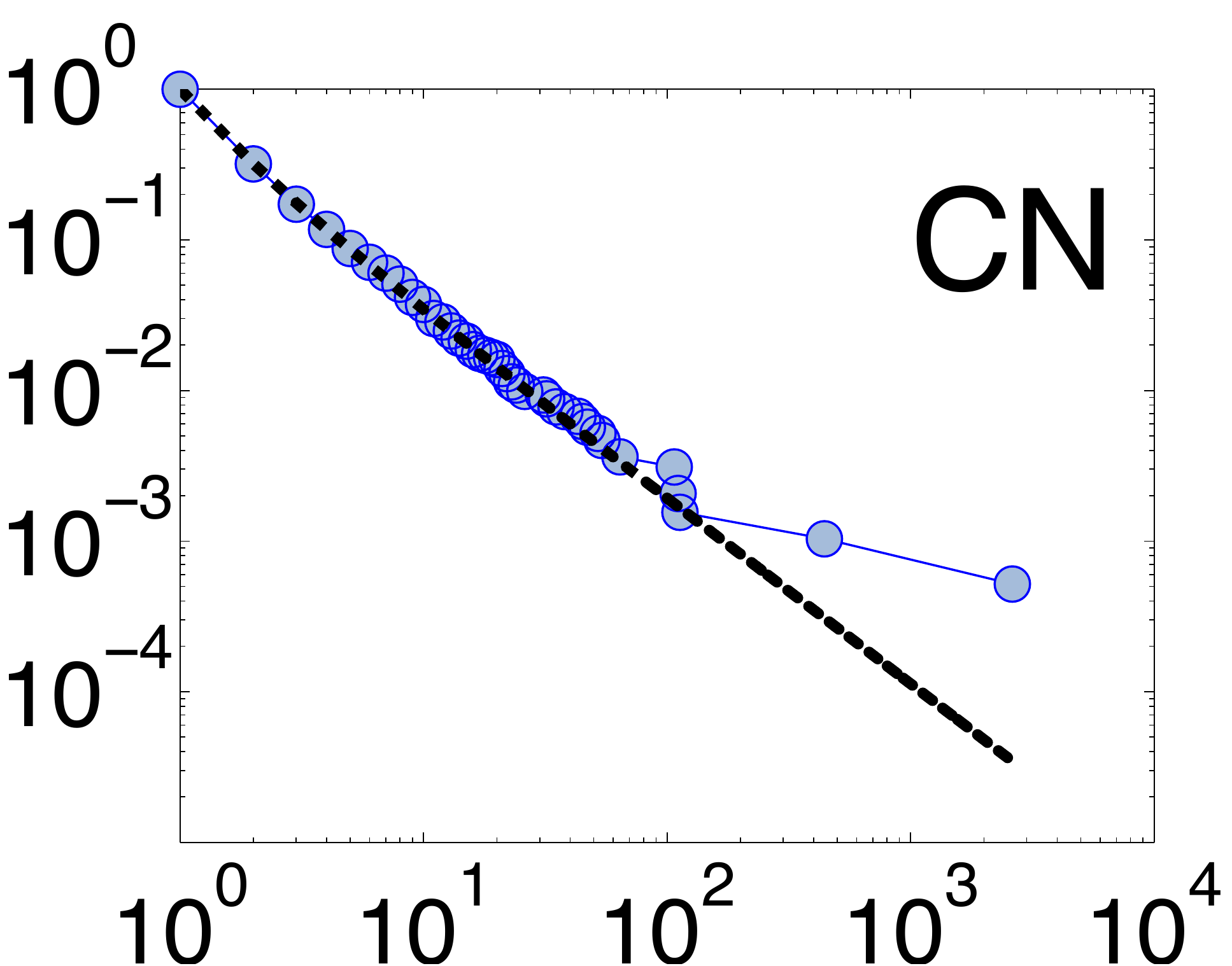}
\includegraphics[width=0.16\textwidth]{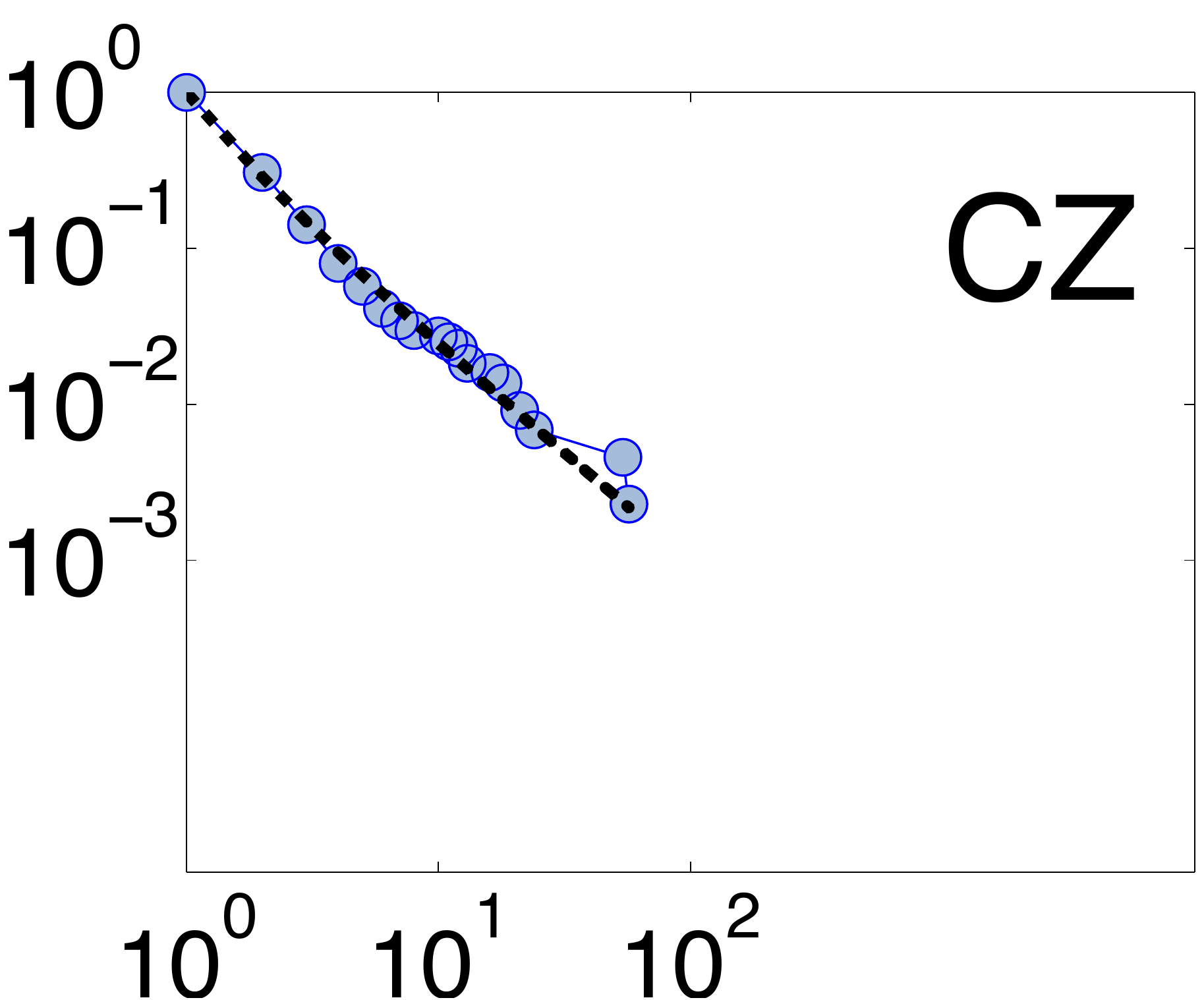}
\includegraphics[width=0.16\textwidth]{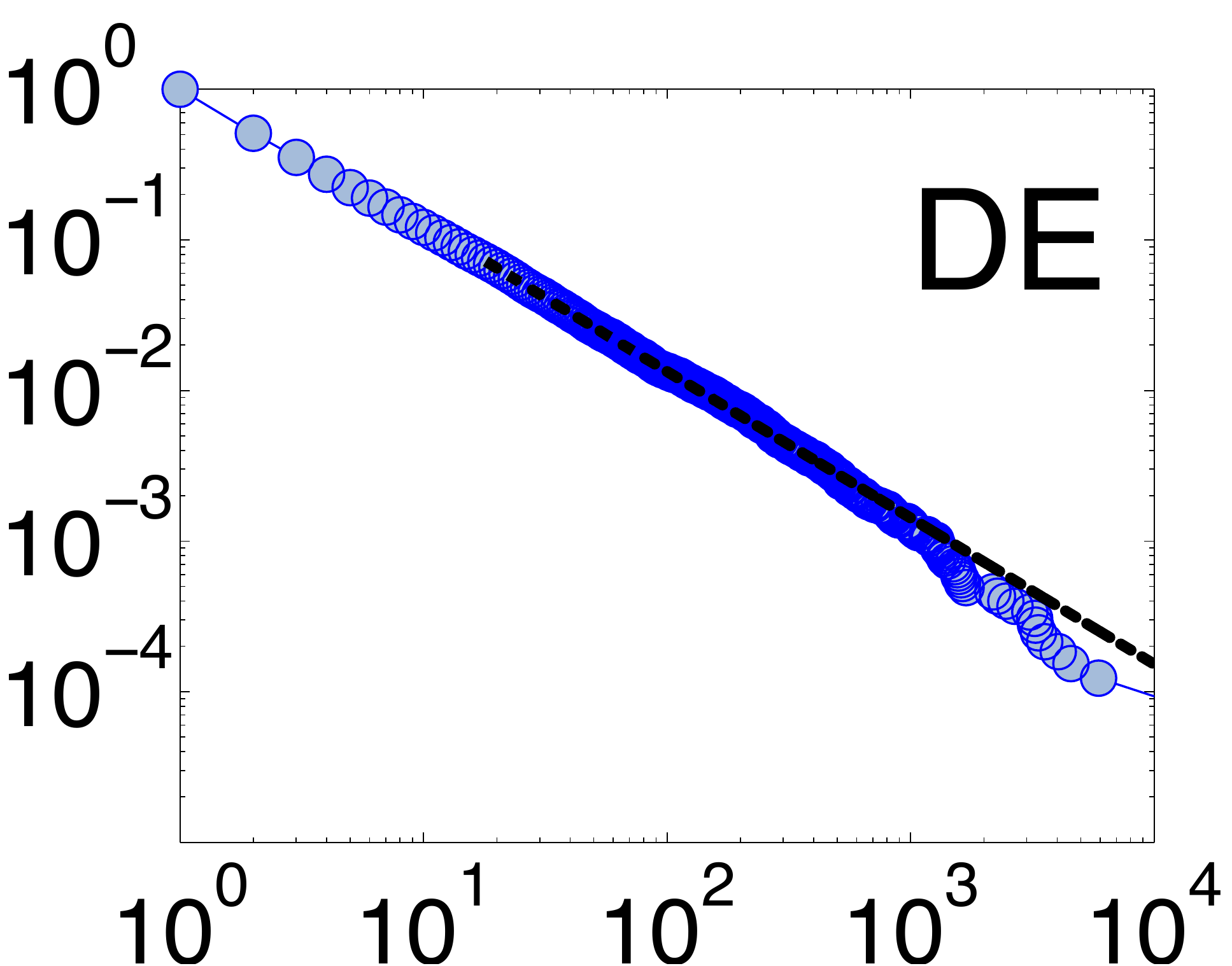}
\includegraphics[width=0.16\textwidth]{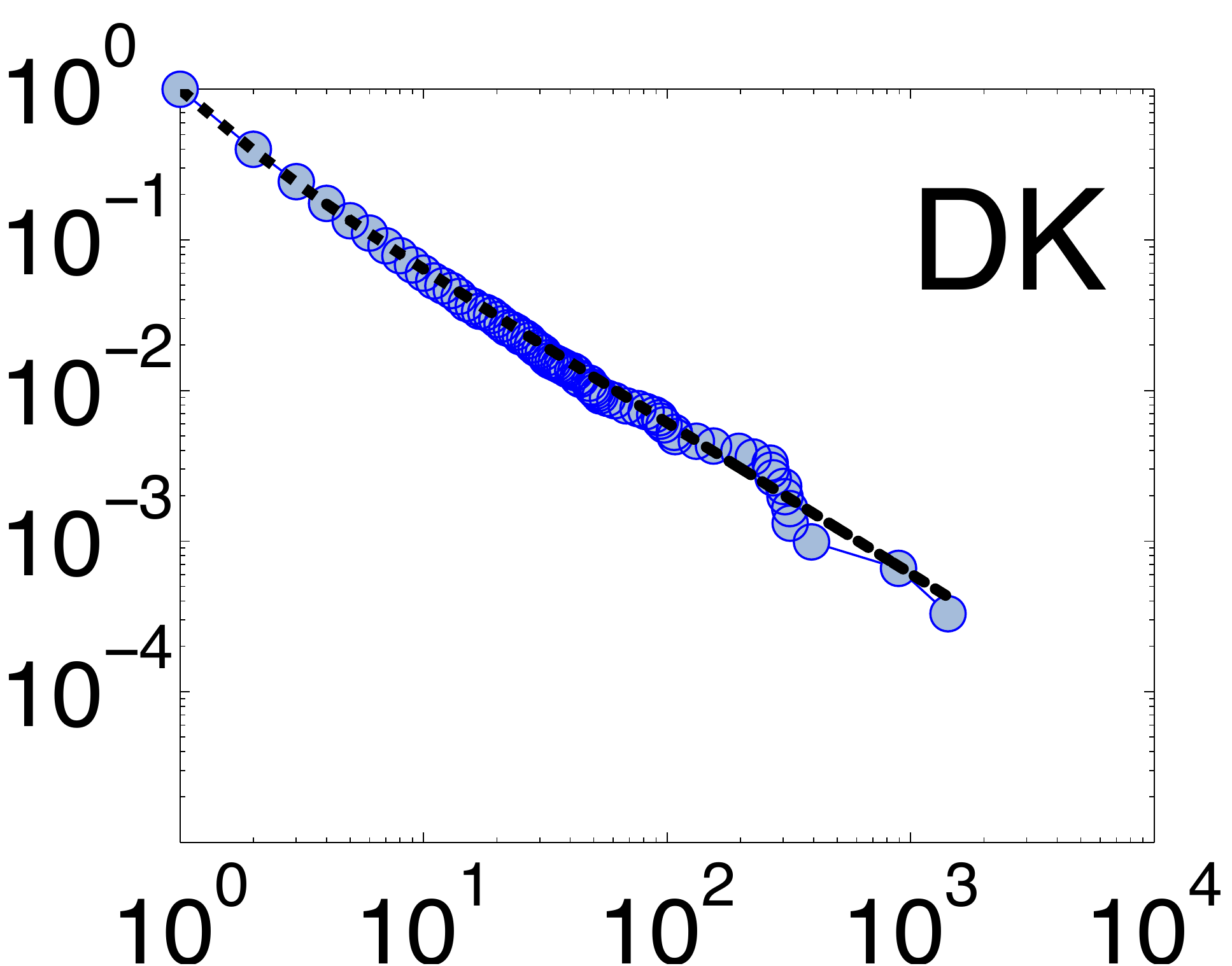}
\includegraphics[width=0.16\textwidth]{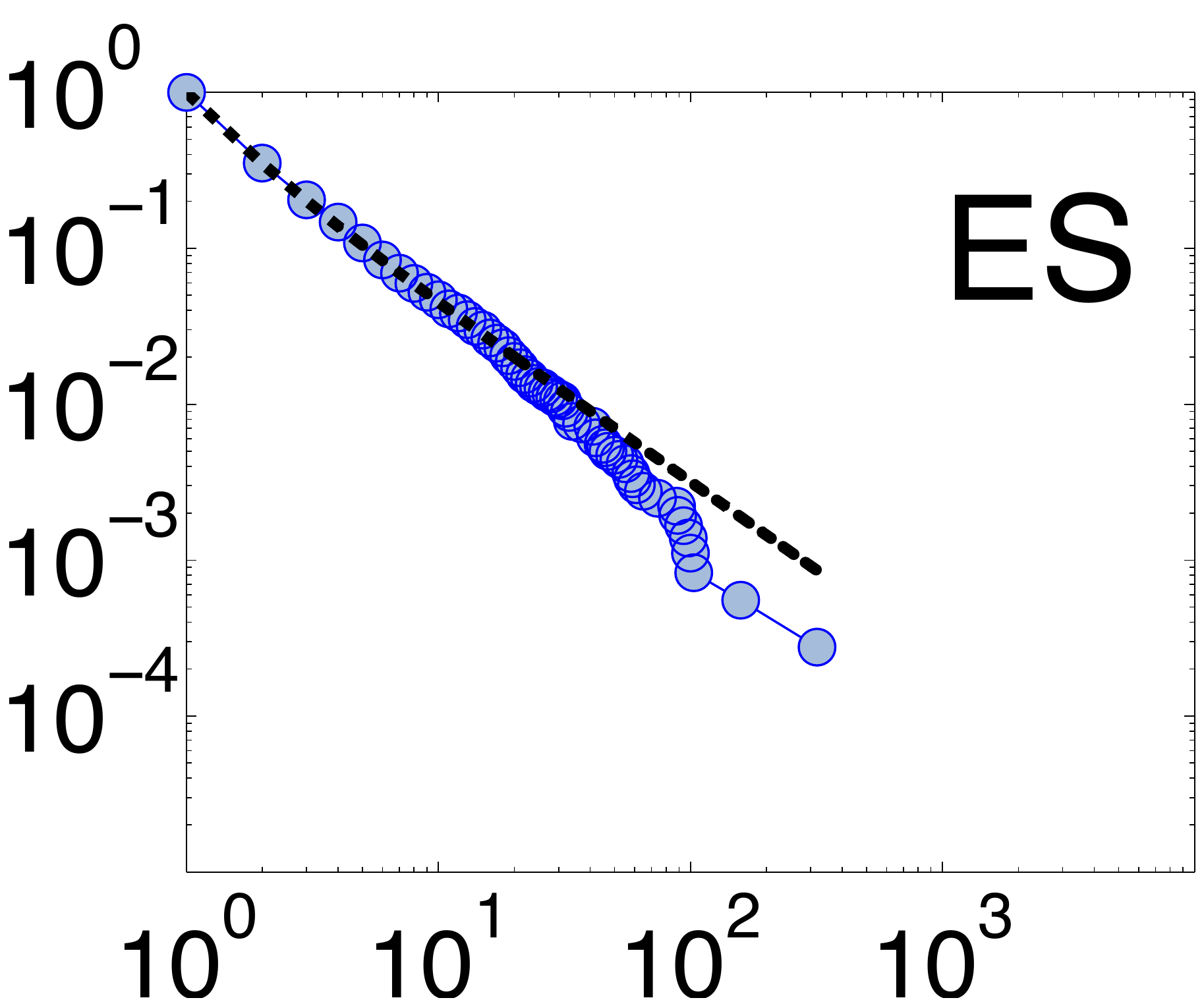}
\includegraphics[width=0.16\textwidth]{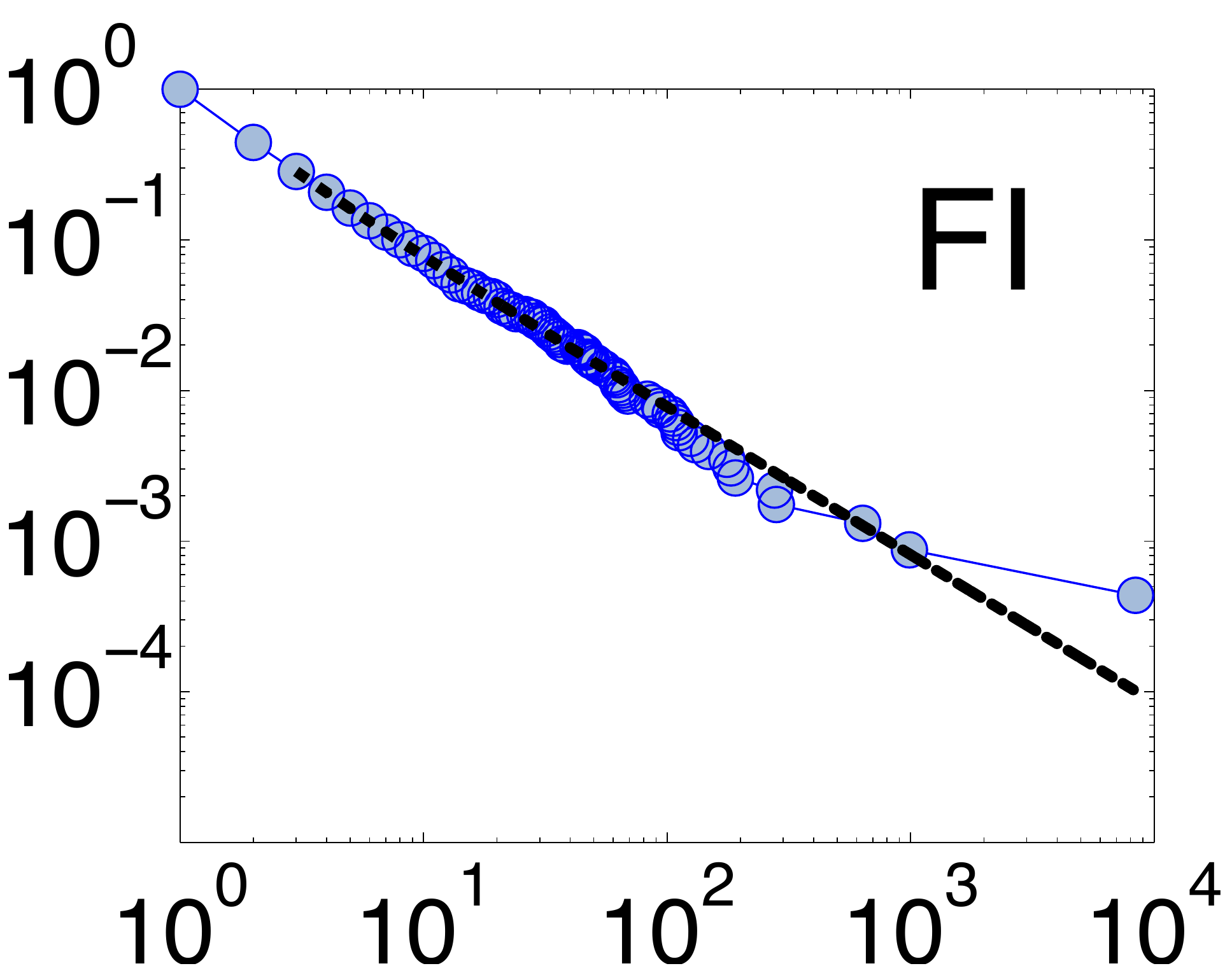}
\includegraphics[width=0.16\textwidth]{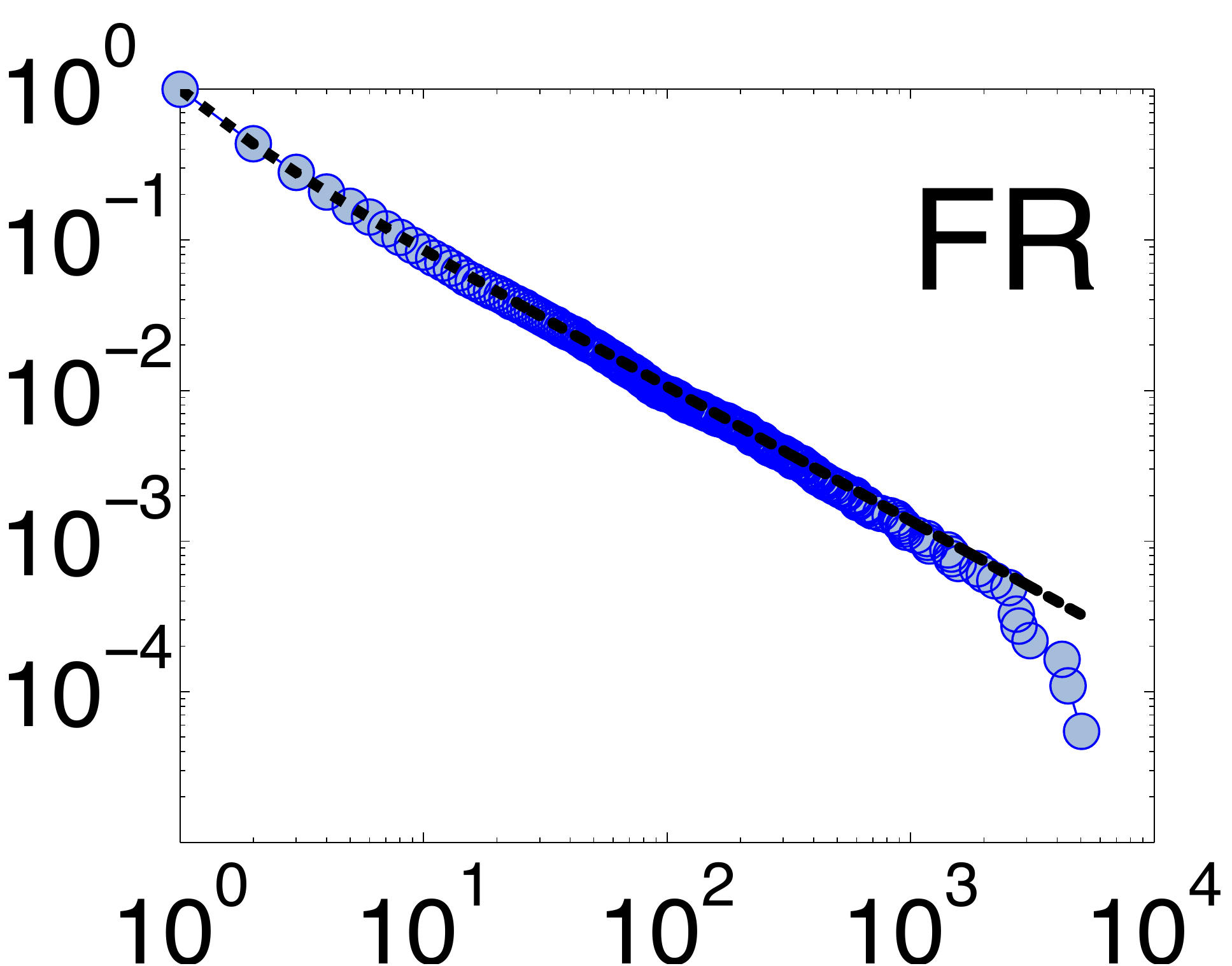}
\includegraphics[width=0.16\textwidth]{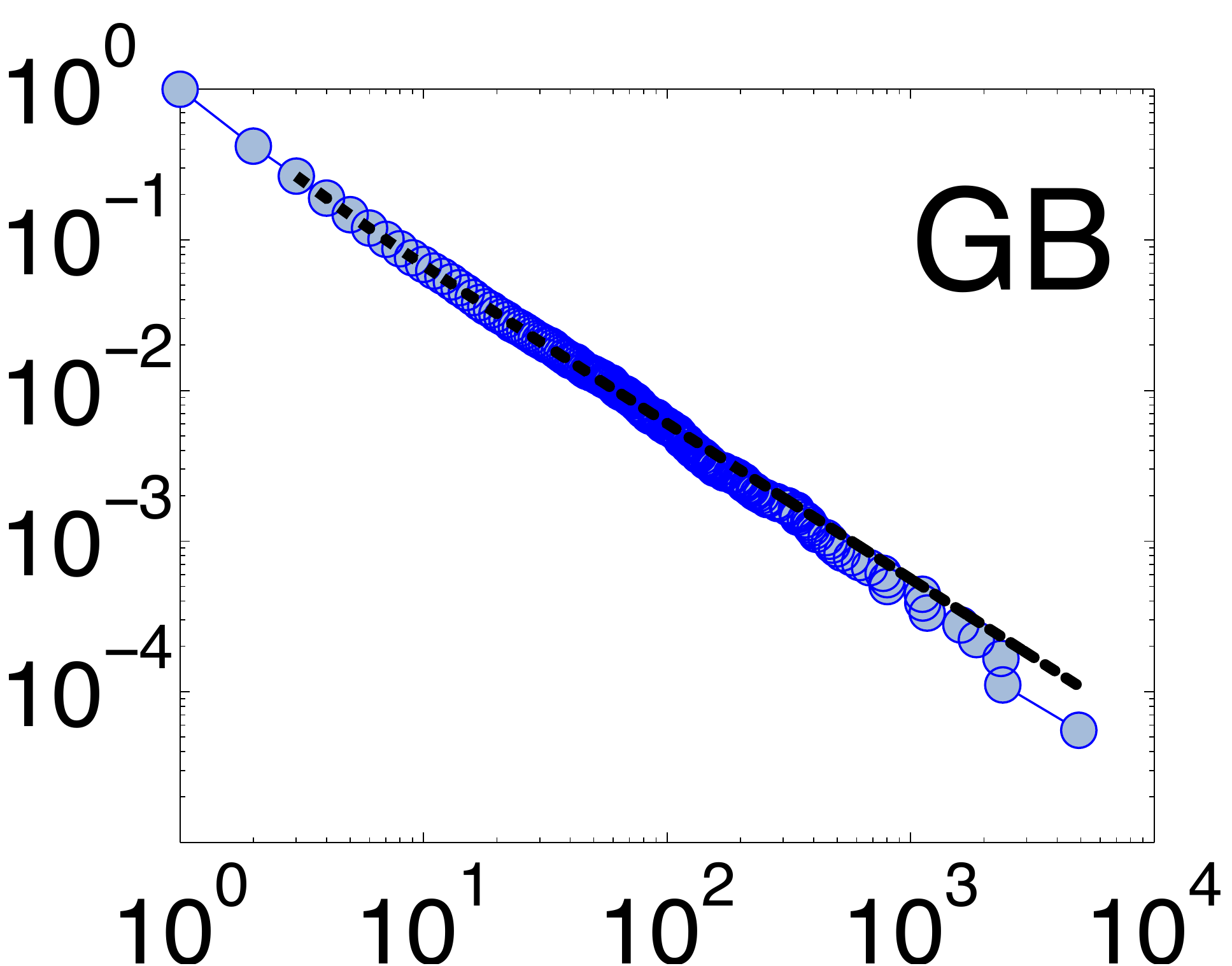}
\includegraphics[width=0.16\textwidth]{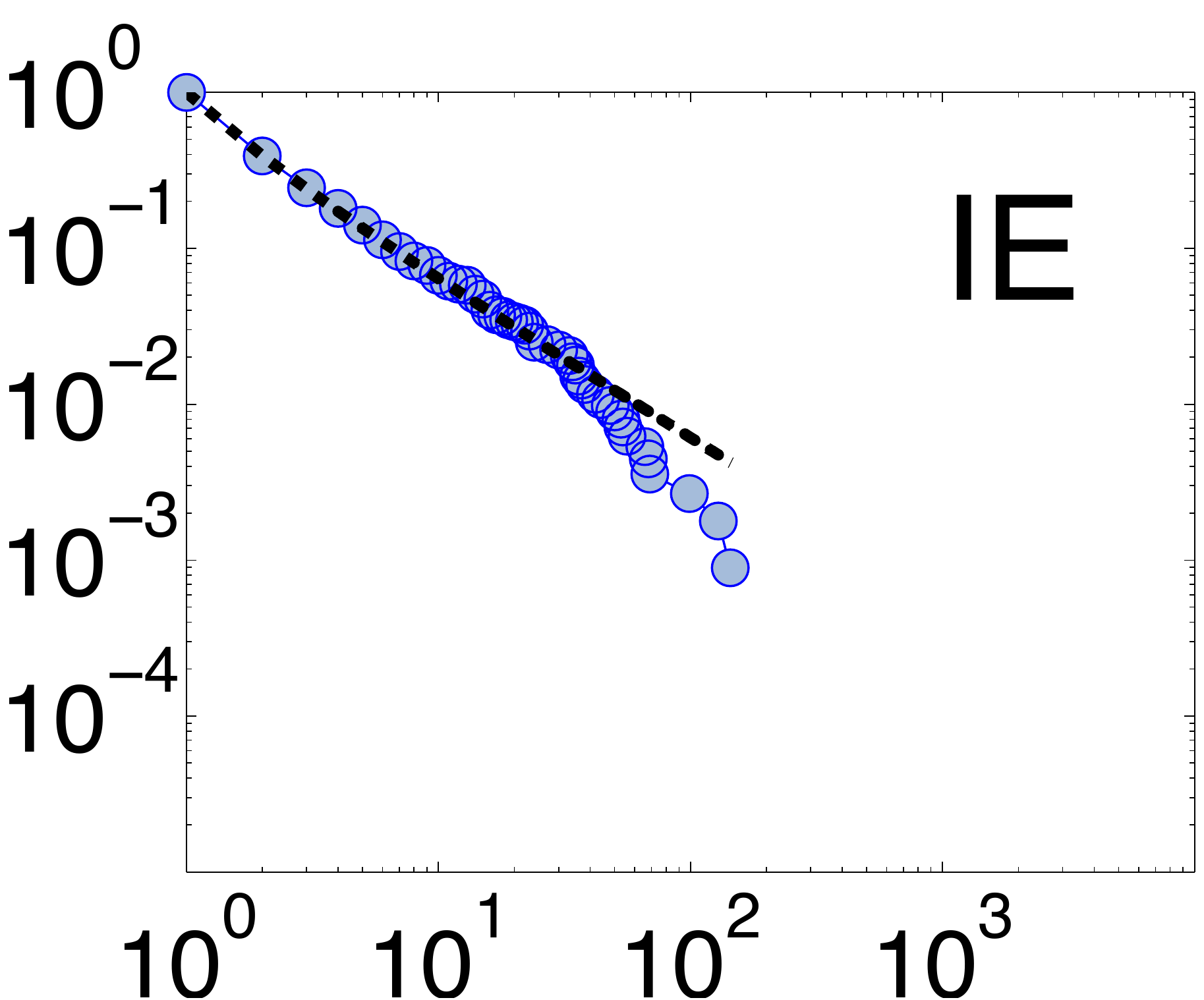}
\includegraphics[width=0.16\textwidth]{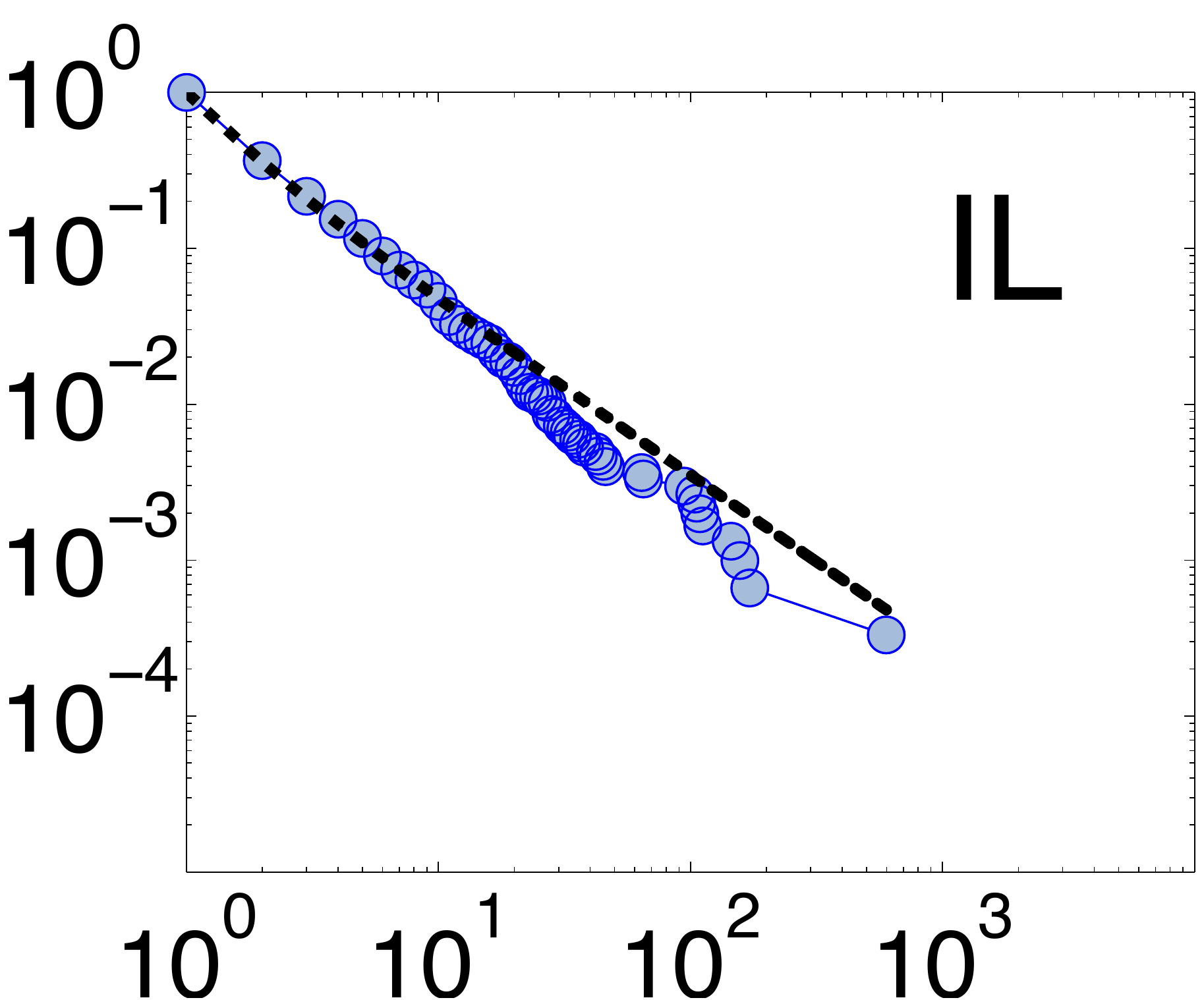}
\includegraphics[width=0.16\textwidth]{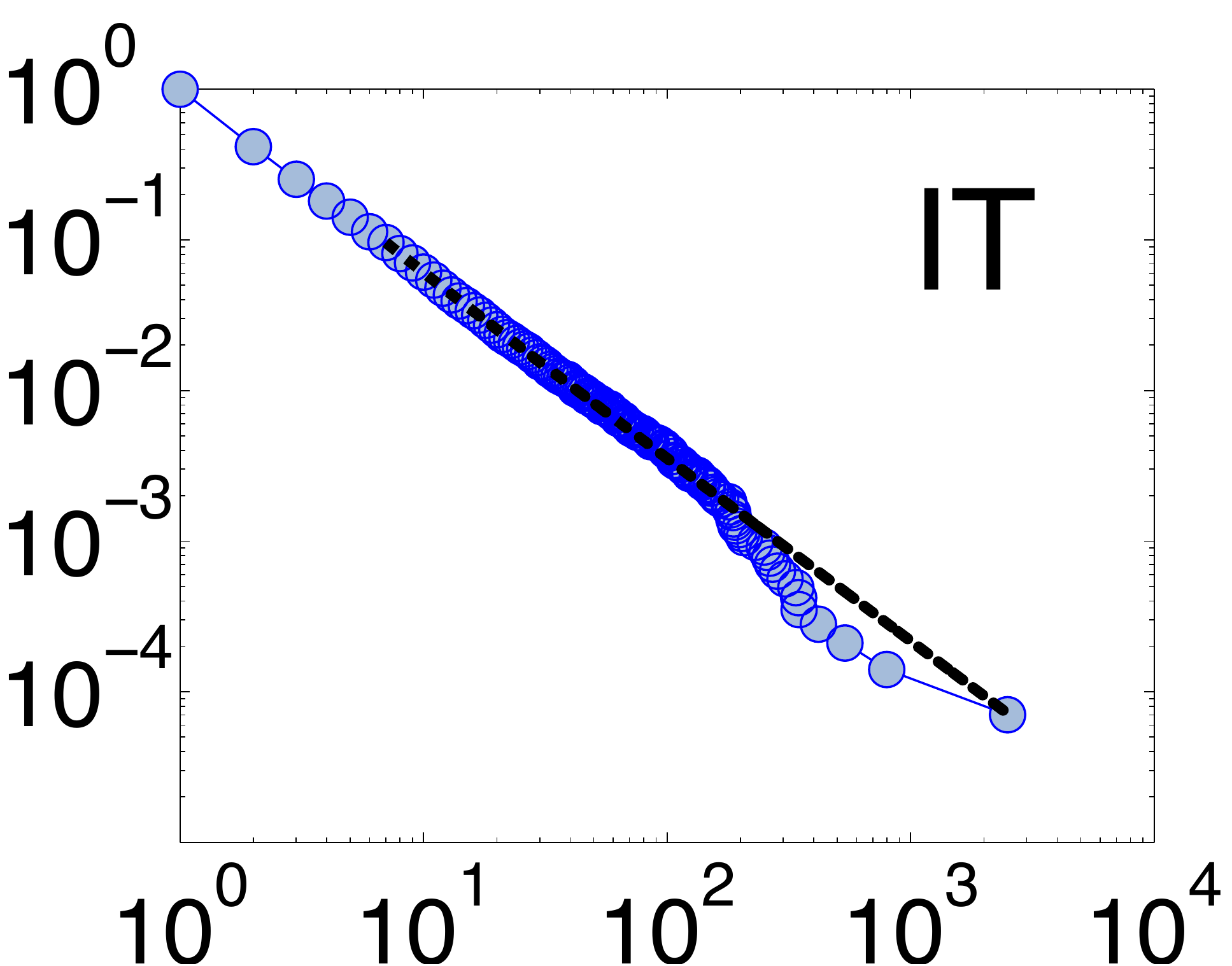}
\includegraphics[width=0.16\textwidth]{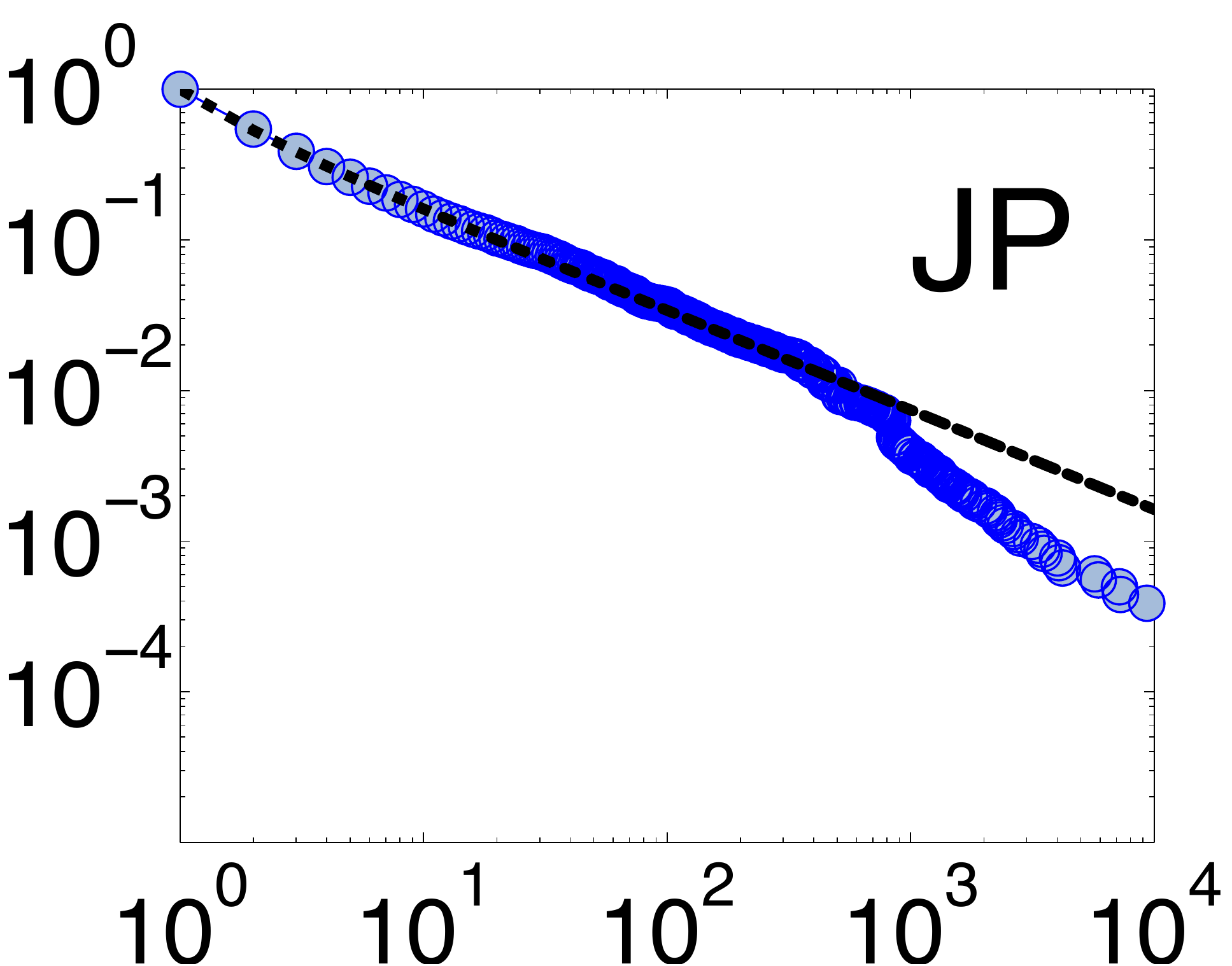}
\includegraphics[width=0.16\textwidth]{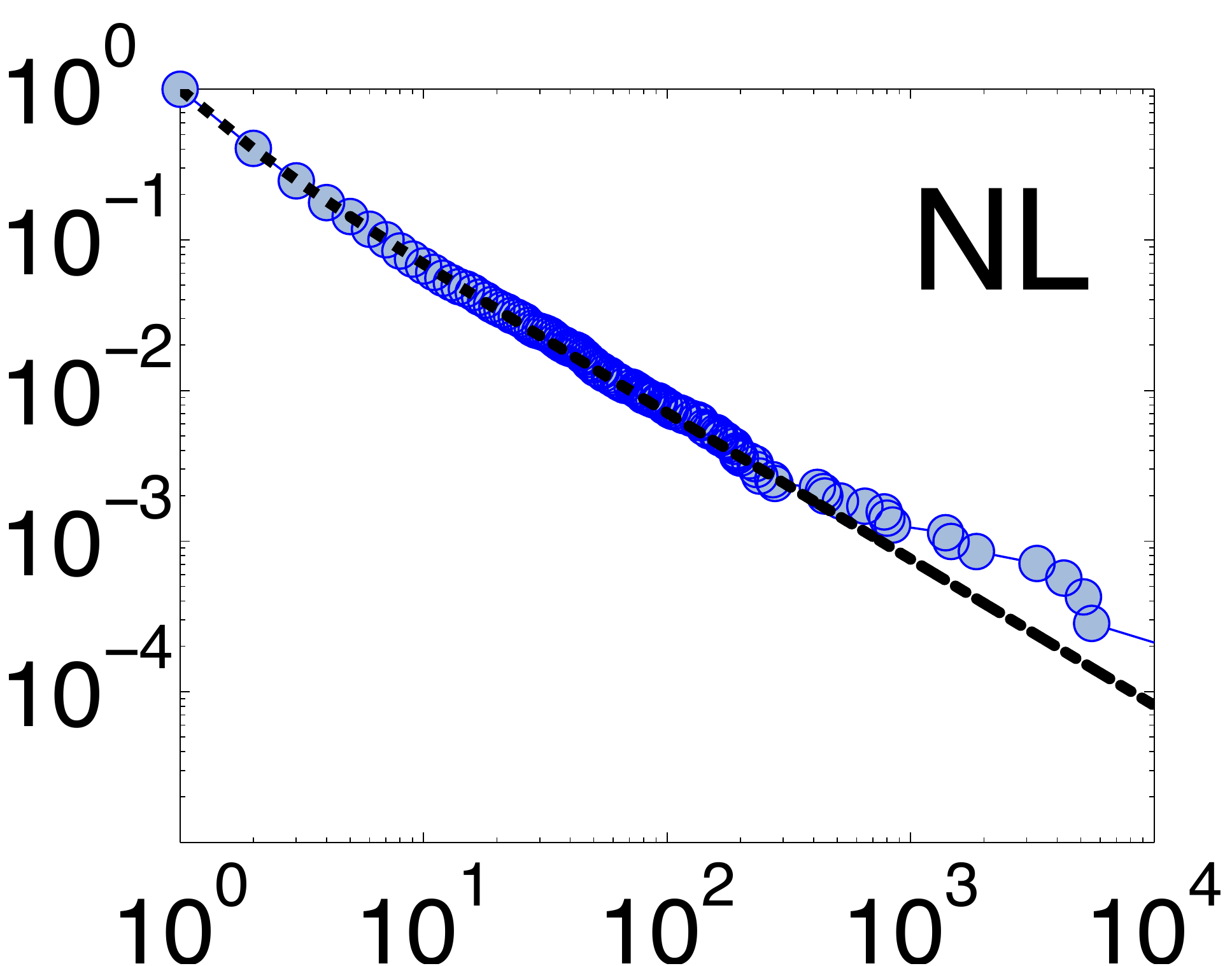}
\includegraphics[width=0.16\textwidth]{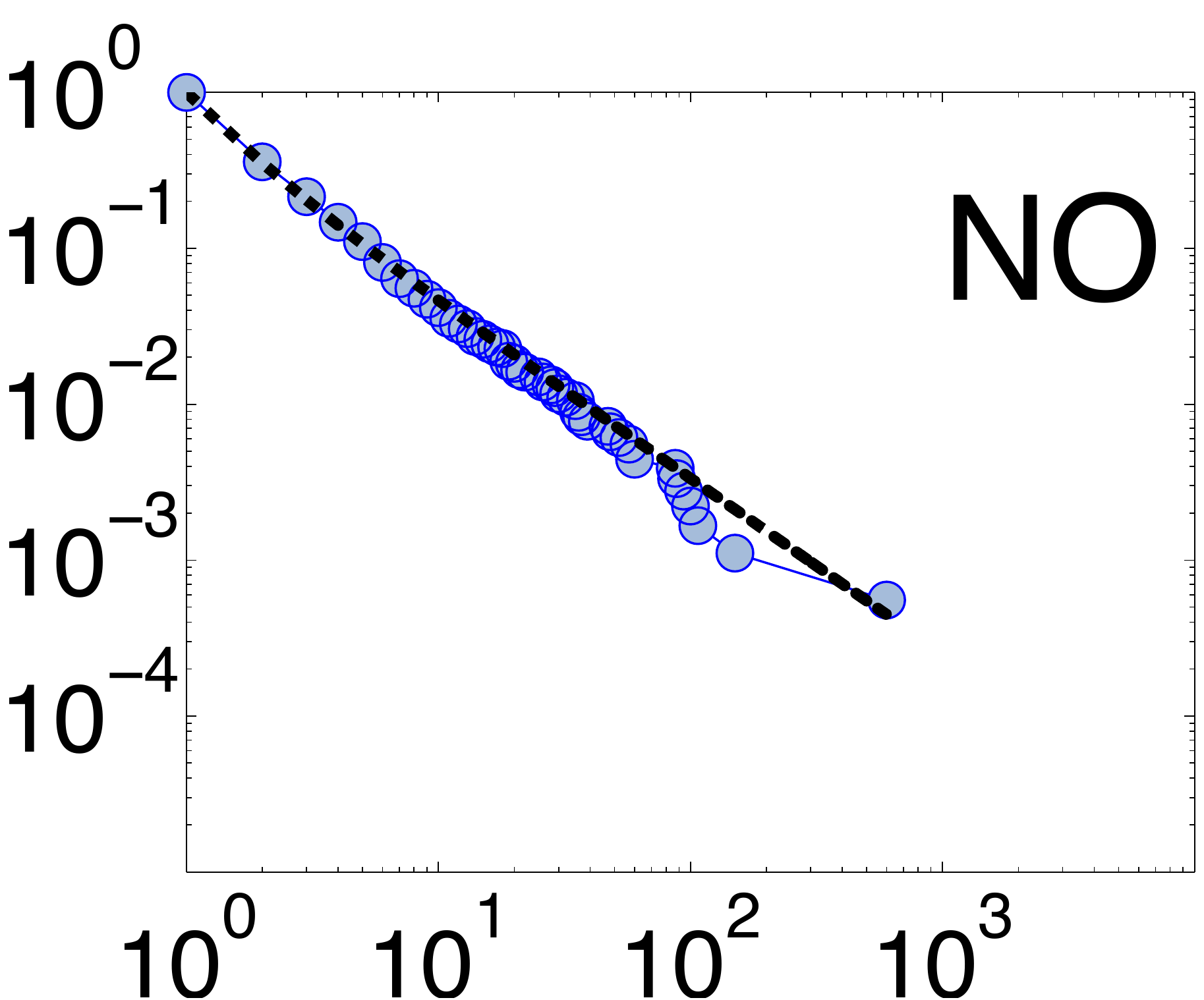}
\includegraphics[width=0.16\textwidth]{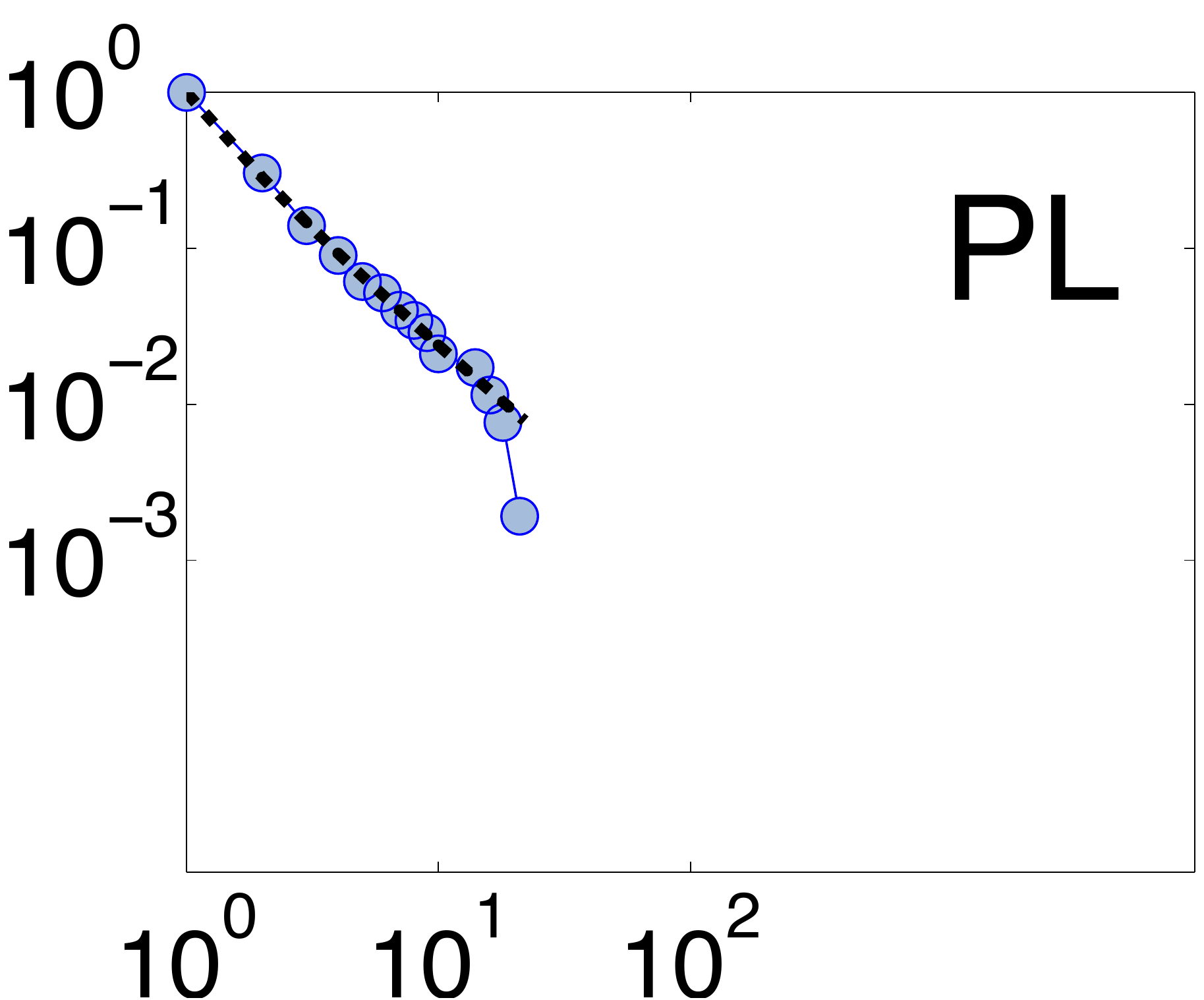}
\includegraphics[width=0.16\textwidth]{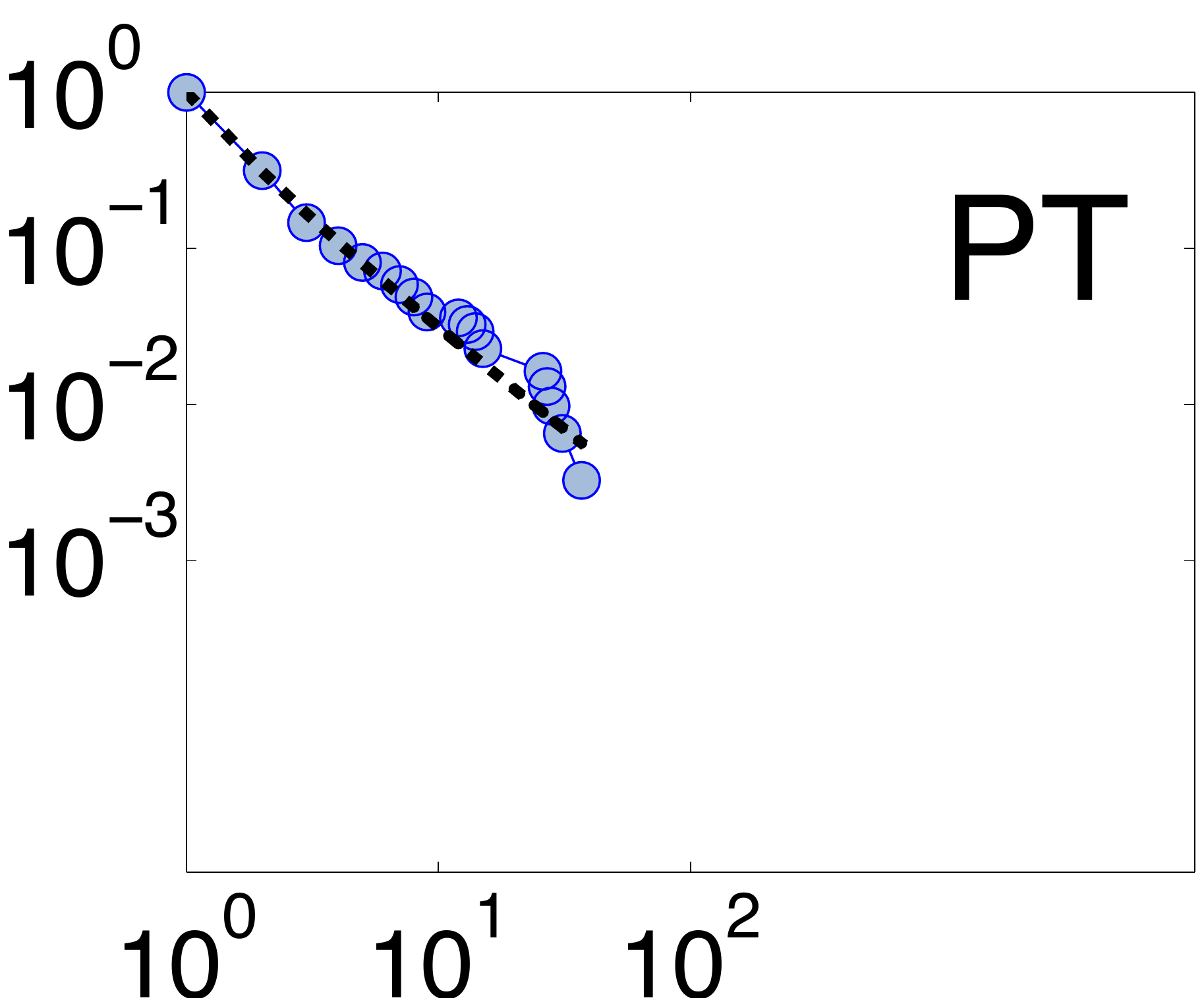}
\includegraphics[width=0.16\textwidth]{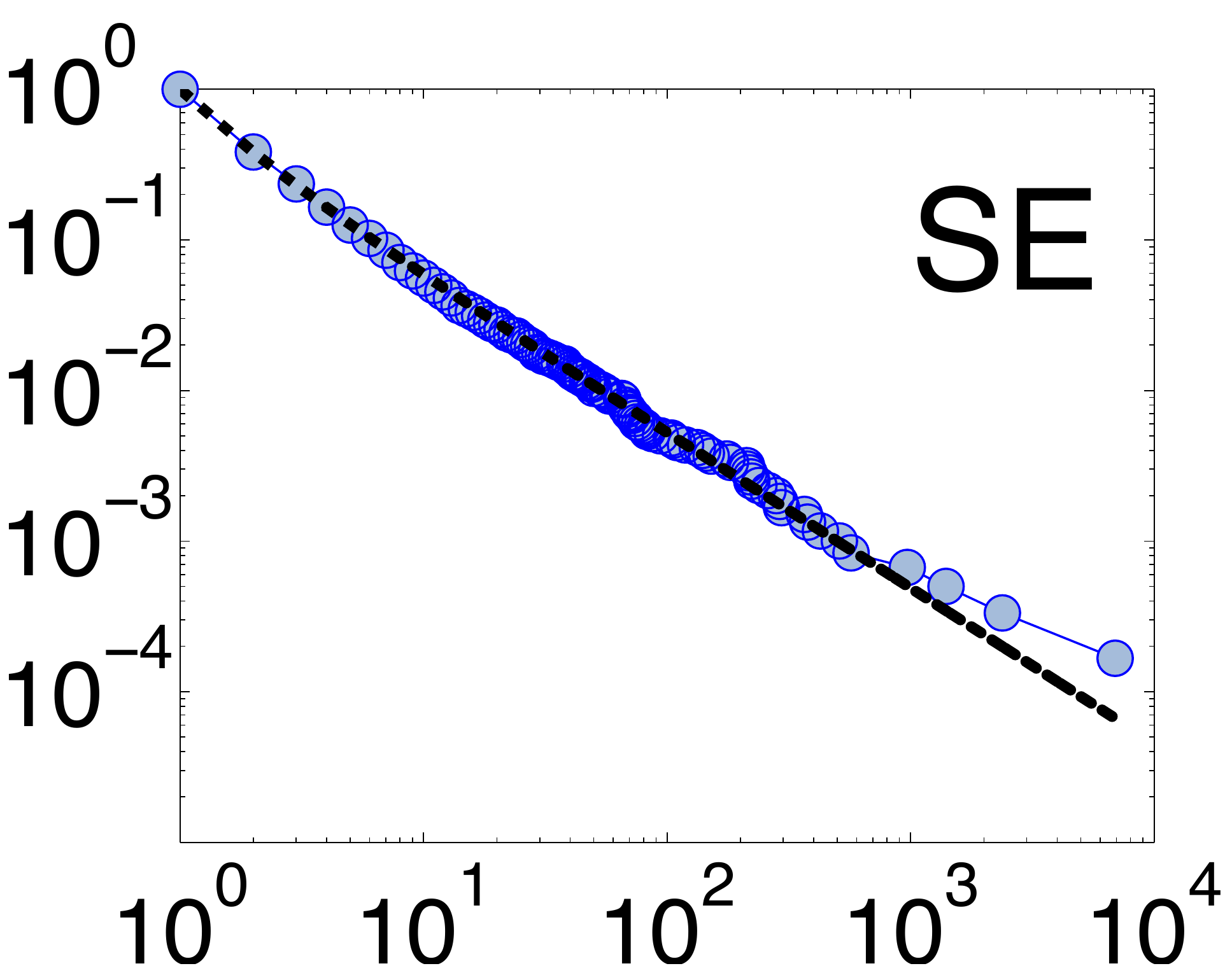}
\includegraphics[width=0.16\textwidth]{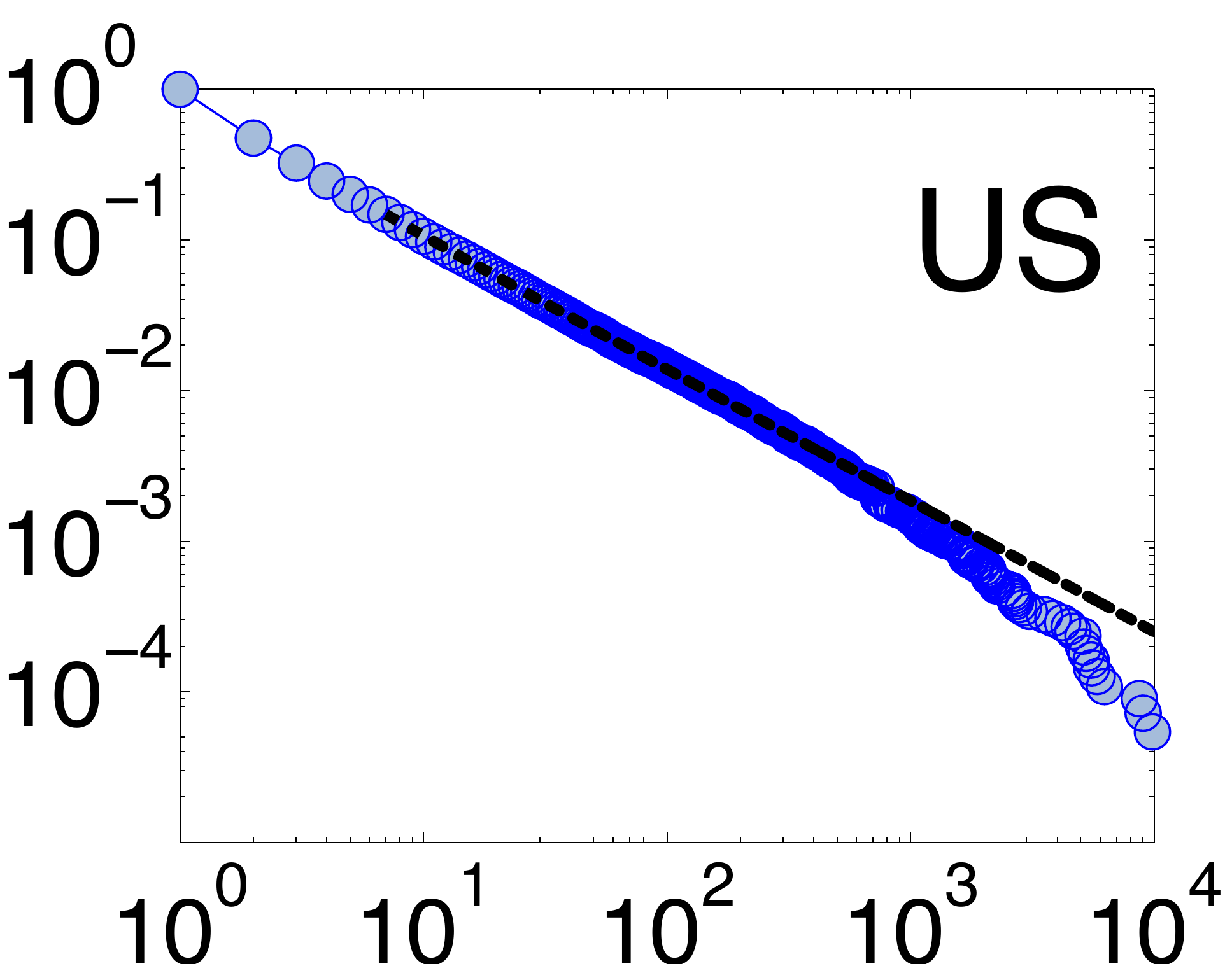}
\\
\caption{\label{fig2}(Color online) CDF plots of EPO patent distributions for all 22 countries, ordered by country code (blue circles) along with a best fit power law model (black dotted line). The match to a power law fit is generally good.}
\end{figure*}

We use data from the OECD HAN data set from July 2011. This includes patent applications filed at the European Patent Office (EPO) from 1977--2007 with partial data from 2007 onwards. The data set covers 22 countries and has been harmonised to correct for cases where applicant details have been recorded differently on different patent applications. The use of the harmonised data is important. When the same analysis was performed with non-harmonised data, large errors resulted from fragmentation where a single unique applicant appeared as several different applicants due to, for example, variations in recording the applicant name.

Power law distributions are only one of many {\it right-skewed}, or {\it heavy-tailed} distributions. By power law distribution, we mean one where the probability distribution, $P[X=x]$ satisfies
$$
P[X=x] = C x^{-\alpha} \mbox{ for } x\geq x_{min}
$$
where $\alpha$ is called the power law exponent or scaling parameter. In the cases we deal with, $x$ is a discrete random variable. To ensure the patent distributions are power laws, we follow the procedure described in \cite{Clauset:2009}. We use a maximum likelihood estimator for a discrete power law distribution to fit the exponent, and estimate  the power law cut-off $x_{min}$~\footnote{For most data, a power law is only fitted to the tail of a distribution, i.e. for values greater than some $x_{min}$. The patent distributions are interesting in that most countries show a good power law fit for the entire range of the data.} by choosing the value which minimises the Kolmogorov--Smirnov (KS) statistic $D=\max_{x\geq x_{min}}|S(x)-P(x)|$, where $S(x)$ is the cumulative density function (CDF) of the data being fitted and $P(x)$ is the CDF of the fitted model distribution. The standard deviations of the fitted values were calculated using a boot-strapping method, drawing a sequence of points $\{\hat{x}_i\in \{x_i\},i=1,\ldots,n\}$ at random, uniformly, and with replacement from the original distribution. The fitted values are listed in Tab.~\ref{tab1} along with their estimated standard deviations. The cumulative density functions for the data of the 22 countries and the fitted power laws are shown in Fig.~\ref{fig2}; the fit is generally strong, with only small deviations between the data and the fitted models over four orders of magnitude for most countries.

\begin{table}
\caption{\label{tab1}The best fit of the empirical data to a power law model is achieved with a power law exponent $\alpha$ and cut-off $x_{min}$. The estimated standard deviation in these parameters is given by $\sigma(\alpha)$ and $\sigma(x_{min})$ respectively. The values for $p$ indicate the ``goodness of fit'' of the empirical data to a power law model. Also given, is $N_{\mbox{app}}$ the number of applicants and $N_{\mbox{pat}}$ the total number of patents held.}
\begin{ruledtabular}
\begin{tabular}{l|l|l|l|l|l|l|l}
ISO&$\alpha$ & $\sigma(\alpha)$ & $x_{min}$ & $\sigma(x_{min})$ & $p$ & $N_{\mbox{app}}$ & $N_{\mbox{pat}}$ \\\hline
	AT &1.97&0.074827&3&1.0777&0.01&3214&18398\\
	BE &1.93&0.017827&1&0.1&0.24&2746&20470\\
	CA &1.99&0.022143&1&0.3266&0.16&4842&24276\\
	CH &1.86&0.036726&2&1.9159&0.01&7907&74987\\
	CN &2.23&0.031152&1&0.14071&0.81&1930&7879\\
	CZ &2.36&0.12005&1&0.25643&0.16&438&910\\
	DE &1.97&0.059017&18&6.681&0.29&32558&391834\\
	DK &2&0.031161&1&0.39492&0.52&3039&15081\\
	ES &2.13&0.020977&1&0&0.14&3614&10408\\
	FI &1.98&0.053119&3&0.68895&0.36&2289&20378\\
	FR &1.89&0.0096839&1&0.25643&0.45&18317&158608\\
	GB &2.03&0.035597&3&0.98985&0.79&18041&99027\\
	IE &2&0.029475&1&0&0.63&1121&4074\\
	IL &2.11&0.021996&1&0&0.04&3019&9138\\
	IT &2.21&0.0996&7&2.9426&0.99&14255&57260\\
	JP &1.66&0.006795&1&0.17145&0&18121&508774\\
	NL &1.97&0.012978&1&0.2&0.43&7043&79976\\
	NO &2.12&0.027157&1&0.1&0.28&1803&5742\\
	PL &2.37&0.13149&1&0.37753&0.15&523&985\\
	PT &2.27&0.080284&1&0.1&0.29&308&707\\
	SE &2.03&0.015404&1&0&0.42&5998&35655\\
	US &1.87&0.024768&7&2.2293&0.61&55539&654304\\
\end{tabular}
\end{ruledtabular}
\end{table}

The values in Tab.~\ref{tab1} for the standard errors in $\alpha$ and $x_{min}$ give us an indication of how precise the estimates of the best fit parameters are, but they do not tell us whether the power law model itself is a good fit. The CDF plots for Austria (AT), Japan (JP), and Switzerland (CH) show significant weight in the portion of their distributions where they deviate from the best fit line. This suggests that a power law model is a poor fit for these countries, despite the small values of $\sigma(\alpha)$ and $\sigma(x_{min})$.

To quantify the goodness-of-fit of a power law model to the empirical data we calculate a so-called $p$ value \cite{Clauset:2009}. The value of $p$ is essentially the fraction of the time when we might expect a goodness-of-fit as poor, or poorer than, that of the empirical data purely due to statistical fluctuations. A $p$ value of 1 would indicate that the amount of mis-fit between the data and a power law is entirely attributable to statistical fluctuations. Values of $p$ less than a threshold in the range of $0.05$ to $0.1$ are typically used to rule out a power law fit. A threshold of $p\geq0.1$ would exclude only four countries --- Austria, Switzerland, Israel, and Japan. However, the main feature of Tab.~\ref{tab1} is that the power law exponents for the 22 countries differ by an amount greater than their estimated standard deviations. The exponents mostly lie between 2 and 2.5, with the exponent for many countries being close to $\alpha=2$ --- the threshold below which the mean value of the distribution diverges. For countries with $\alpha>2$, the expected mean value for the fitted power law distribution is given by $\langle x \rangle = (\alpha-1)/(\alpha-2)x_{min}$, hence the expected mean number of patent applications per firm is large for many of the countries.  

\begin{figure}
\includegraphics[width=0.95\columnwidth]{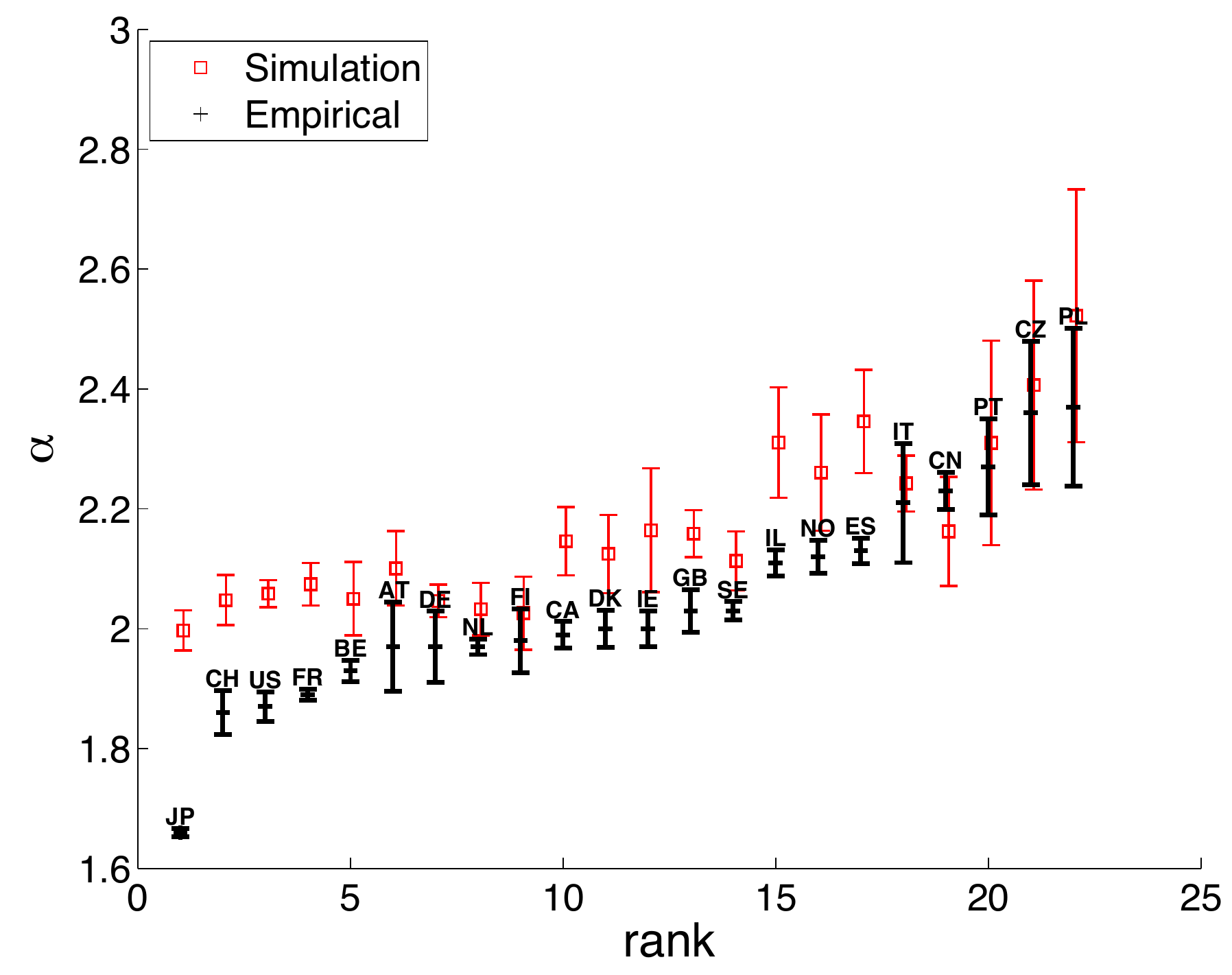}
\caption{\label{fig3}(Color online) Values of $\alpha$, with their associated estimated uncertainties, for the 22 countries in the EPO HAN data set, sorted by $\alpha$, along with average $\alpha_{\mbox{sim}}$ values (and the associated standard deviations) for simulated data. For each country's simulation, the growth rate was determined by the ratio $\gamma=N_{app}/N_{pat}$ from Tab.~ \ref{tab1}. }
\end{figure}

\begin{figure}
\includegraphics[width=0.48\columnwidth]{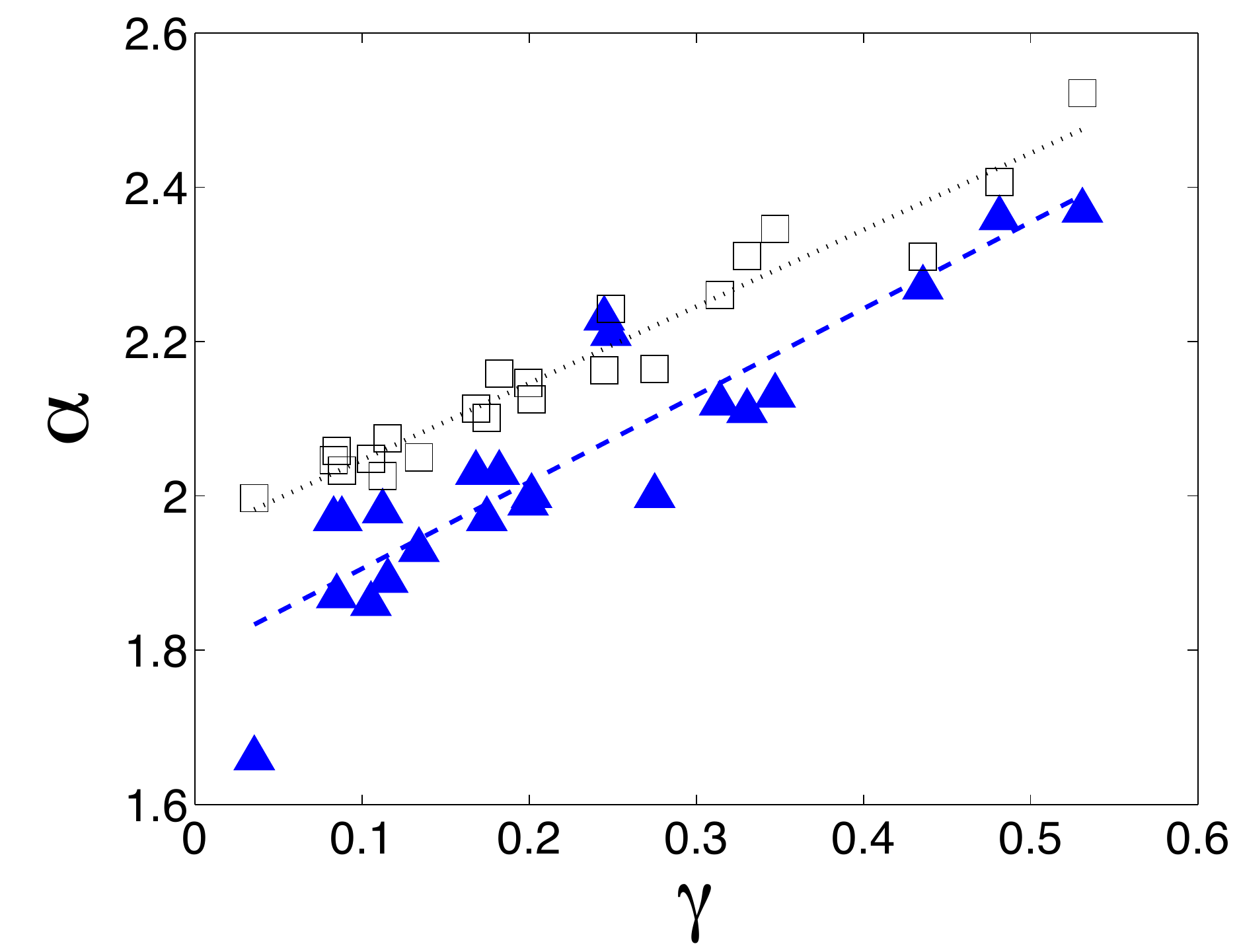}
\includegraphics[width=0.48\columnwidth]{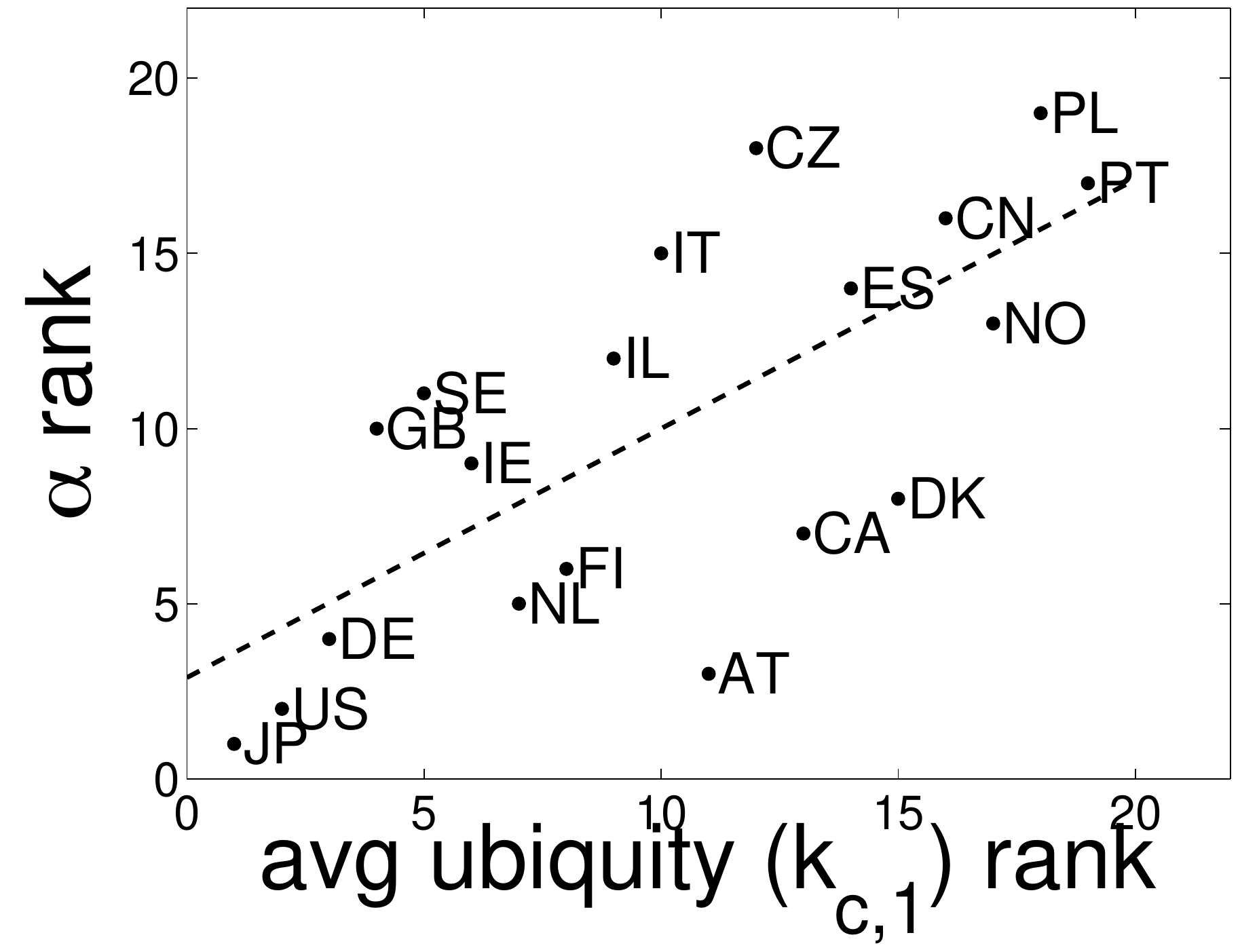}
\caption{\label{fig4} (Color online) Left: Best fit exponents for the empirical (filled triangles) and simulated (empty squares) distributions of each country as a function of $\gamma=N_{app}/N_{pat}$. The linear least square fits are, respectively, $\alpha=1.78+1.12\gamma$ ($R^2=0.81$) and $\alpha_{\mbox{sim}}=1.95+0.99\gamma$ ($R^2=0.94$). Right: Correlation between the country rank of the empirical power law exponent and the country rank based on average export ubiquity ($k_{c,1}$ in \cite{Hidalgo:2009}). The dashed line indicates the linear least-squares fit and has slope 0.71 ($R^2=0.5$).}
\end{figure}

Figure \ref{fig3} shows a plot of the power law exponents and their estimated standard deviations for each country.  We order the countries by the size of the exponent $\alpha$. It is interesting to note that the countries which are often thought of as having innovative or ``high-tech'' economies lie mostly towards the left of the plot, with smaller power law exponents. The link between small exponents and economies with highly specialized firms is reinforced by Fig.~\ref{fig4}(b) where the rank of the countries by exponent is plotted against the rank for the average ``ubiquity'' of the goods exported by that country \cite{Hidalgo:2009} --- a measure of how specialised the goods are.  The correlation between the exponent rank and ubiquity rank indicates that the power law exponents give information about the presence of sophisticated (export) sectors in a country. The lower the value of the power law exponent of a country, the more likely that the country exports a number of specialised goods, exported by few other countries.

Having determined that the distribution of patents amongst applicants follows a power law in many OECD countries, and that the power law exponent differs between countries, we now describe a mathematical model which captures this behaviour. Since it is not clear why the distribution of patents amongst applicants should {\it necessarily} follow a power law, and since the underlying rules or patterns which lead to such a distribution for patents is not obvious, it is important that our generative model follows some set of rules or procedures which could credibly apply to growth in the number of patents.

There is a large literature on generative models for power law distributions, going back almost a century \cite{Mitzenmacher:2004}. 
We use a model equivalent to the Yule process \cite{Yule:1925}, based on two assumptions; 1) growth --- the number of applicants with patents increases over time, and 2) preferential attachment --- the likelihood of an existing applicant acquiring a new patent is proportional to the number of patents that the applicant already holds.  

These assumptions lead to the following algorithm: Beginning with a single applicant holding a single patent, at each time step we either add a new applicant holding a single patent or add a new patent to an existing applicant. The rate at which new applicants are added is determined by the growth rate $\gamma$, fixed throughout the simulation. When a patent is added to an existing applicant the probability that it is attached to applicant $i$ is given by $k_i/\sum_{j=1}^N k_j$, where $k_j$ is the number of patents held by applicant $j$ and $N$ is the total number of applicants in the model at that time step. Since only a single patent is added at each time step, the (inverse of the) growth rate $\gamma$ gives (for large $N$) the average number of patents per applicant. Since it is desirable that that a model should reproduce known quantities, such as the average number of patents per applicant, we choose $\gamma=N_{\mbox{app}}/N_{\mbox{pat}}$ using the data in Tab.~\ref{tab1}, which ensures that this quantity matches that observed empirically, and eliminates the only free parameter in the model.  

It is not difficult to prove that such an algorithm produces data with a power law tail when the number of steps taken becomes large \cite{Mitzenmacher:2004} and that the power law exponent for the simulated distribution tends towards $\alpha=2+\gamma$ \cite{Newman:2005}. Hence, the exponents of the simulated distributions are bounded below by 2, approaching this limit as the average number of patents per applicant becomes large. 

Using this algorithm we simulated the growth of the corresponding distribution of patents 500 times for each country, determining the average value of $\alpha$ and its corresponding standard deviation.  Fig.~\ref{fig3} shows the values of $\alpha$ and the estimated standard errors for both the simulated and empirical data sets, again ordered by $\alpha$ (for the empirical data). While there are as many countries for which the error bars of the simulated and empirical data do not over lap as those for which they do, there is a clear qualitative fit: countries with lower values of $\alpha$ in the empirical data, show the same pattern in the simulated data. In Fig.~\ref{fig4}(a) we show the relationship between the exponent $\alpha$ and the growth rate $\gamma$ for empirical and simulated data. Both show a clear correlation although the empirical data consistently has lower exponents than the simulated data. The linear regression fit for the simulated data is very close to the asymptotically expected results, so the gap between the simulated and empirical exponents is not due to the finite duration of the simulations.

The relatively good agreement between the simulated and empirical data gives support to the assumption that applicants holding many patents are more likely to acquire further patents --- future innovators are likely to also be past innovators.  However we note that in our model, applicants can continue to acquire patents indefinitely. Clearly, this assumption is not realistic --- applicant firms can go out of business or be acquired by other firms.  Similarly, a single patent may be owned by more than one applicant~\footnote{Although 88\% of the patents in the HAN EPO data set have a single applicant.} resulting in a network of co-applicants. Both these effects will alter the patent distribution from that seen in the simulations.

We also investigate the relationship between the patent distributions and national expenditures on R\&D. In an innovation ecosystem gross expenditure on research and development (GERD) and business expenditure on research and development (BERD) might be thought of playing a role similar to that played by biomass in ecosystems \cite{Oecd:2006}. Just as the total biomass of an ecosystem can be used to normalise frequency versus body size distributions such that ecosystems of different sizes can be direfectly compared; rescaling the absolute number of patents for a country by that country's absolute GERD or BERD causes the patent distributions to collapse on one another (not shown here). However, it is also interesting to see how the exponent $\alpha$ is related to expenditure on R\&D. 

\begin{figure}
\includegraphics[width=0.45\columnwidth]{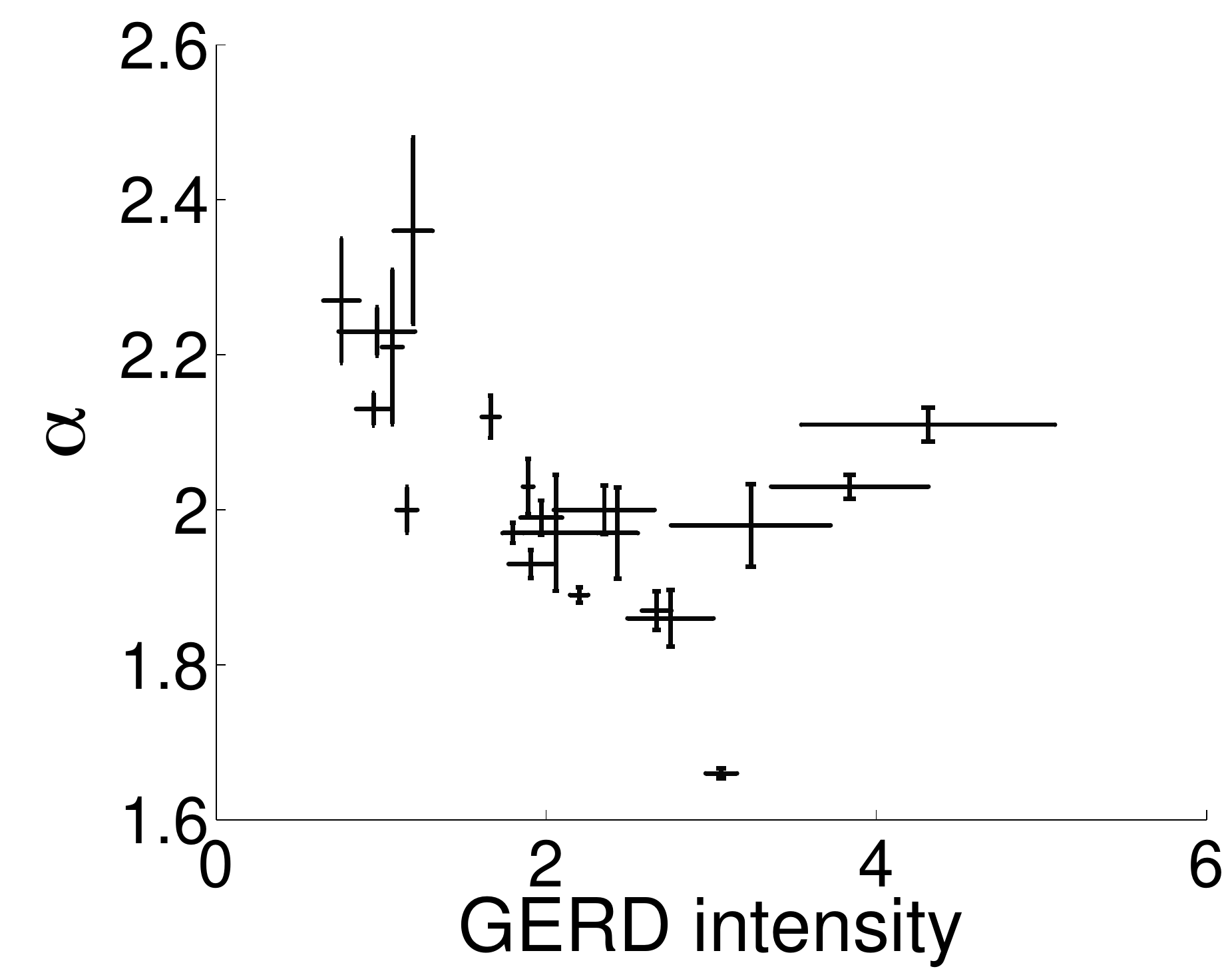}
\includegraphics[width=0.45\columnwidth]{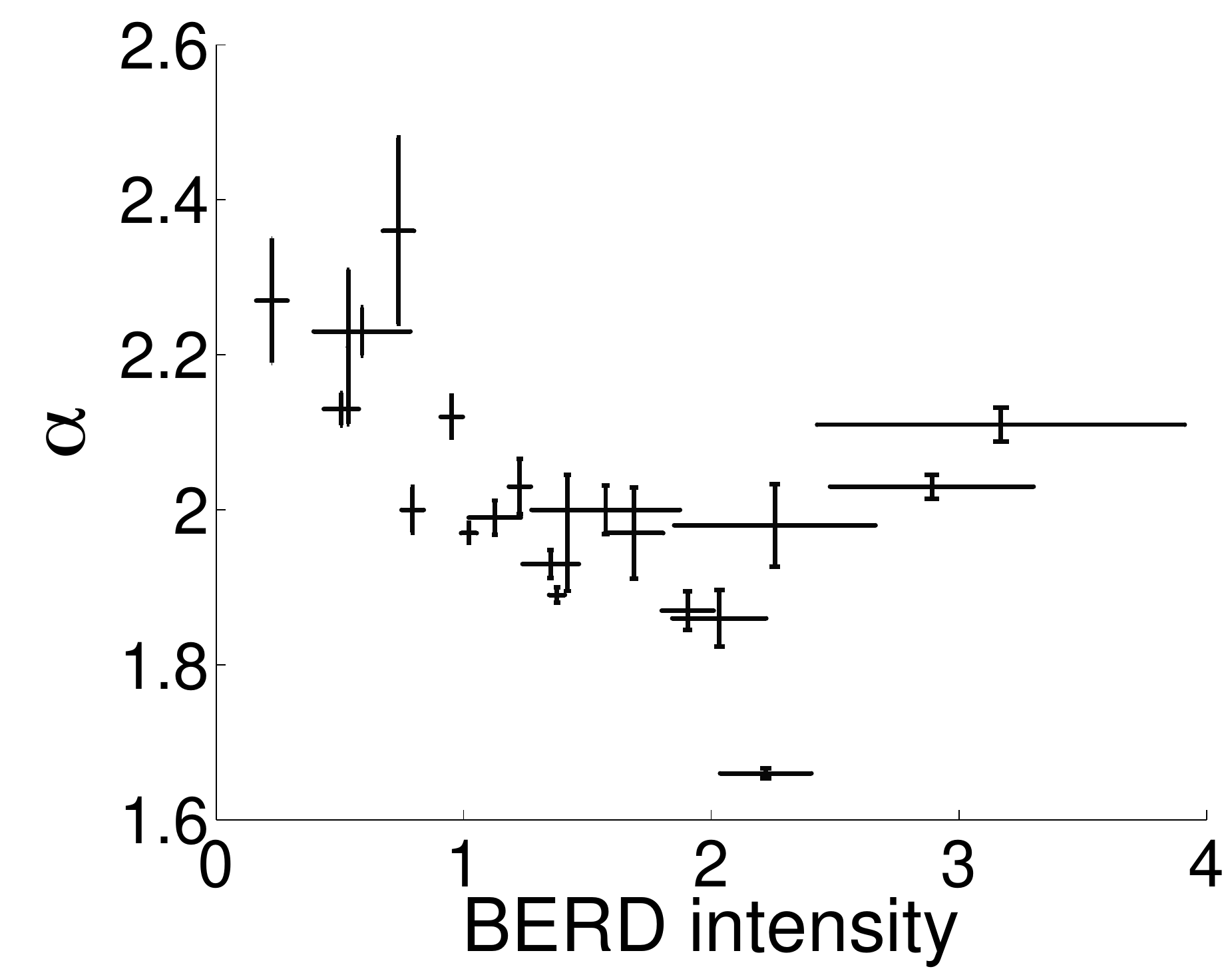}
\caption{\label{fig5} Power law exponents for EPO patent distributions of 22 OECD countries versus GERD (left) and BERD (right) intensity --- gross, resp., business expenditure on R\&D as a percentage of GDP.
Vertical bars indicate the estimated standard error in the $\alpha$ values, horizontal bars indicate the standard deviation in the time averaged (1995-2006) OECD data.}
\end{figure}

To investigate this, we plot the EPO power law exponents for the countries in the OECD HAN data set against GERD and BERD intensity as a percentage of gross domestic product (GDP): Fig.~\ref{fig5}. We see a strong correlation between increasing intensity of expenditure on research and development, and lower values of the power law exponent (corresponding to more innovative and more sophisticated economies). An interesting feature is that the decrease in $\alpha$ appears to saturate at about 3\% GERD intensity, or 2\% BERD intensity. Beyond this level of R\&D expenditure, there is no evidence of further flattening of the patent distributions. It is also interesting to note that the correlation of $\alpha$ with both GERD and BERD is the same --- a translation of BERD intensity by around 1\% almost exactly matches the pattern for GERD intensity, implying that both BERD and GERD play similar roles.

In conclusion, we have found that distribution of patents amongst applicants within OECD countries generally follow power laws. This provides a new way of looking at the structure of national economies and strengthens the analogy between innovating firms and ecosystems. Indeed, we find that the power law exponents that describe these distributions differ between countries and are correlated with measures such as national expenditure on research and development, and the ubiquity, or degree of specialisation, of the basket of goods that a country exports. Countries that export more specialised goods tend to have a smaller proportion of companies that hold a larger share of the patents where countries that export more ubiquitous goods tend to have a larger share of patents held in small portfolios. It seems that the innovator of today is more likely to work in the research laboratory of a large multinational company than in the suburban garage or small start-up company.

\begin{acknowledgments}
The authors acknowledge the use of the OECD HAN database, July 2011 and the Matlab code for power law fitting and analysis from Clauset {\it et al} \cite{Clauset:2009}. They wish to thank Catriona Sissons for useful discussions.
\end{acknowledgments}

\bibliography{pl-dist}

\end{document}